\documentclass[pra,aps,superscriptaddress,twocolumn,nopacs,nofootinbib]{revtex4}
\usepackage{graphicx,color,epstopdf}
\usepackage{amsmath,amsfonts,enumerate,amsthm,amssymb,mathtools}
\usepackage{thmtools,thm-restate}
\usepackage{bbold}
\usepackage{hyperref}

\hypersetup{
   colorlinks=true,         
   linkcolor=blue,          
   citecolor=blue,          
   urlcolor=blue            
}

\def\E{ {\cal E} }
\def\Sc{ {\cal S} }
\def\T{ {\cal T} }
\def\P{ {\cal P} }
\def\>{\rangle}
\def\<{\langle}

\newcommand{\ketbra}[2]{\ensuremath{\left|#1\right\rangle\!\!\left\langle#2\right|}}
\newcommand{\matrixel}[3]{\ensuremath{\left\langle #1 \vphantom{#2#3} \right| #2 \left| #3 \vphantom{#1#2} \right\rangle}}
\newcommand{\tr}[2]{\mathrm{Tr}_{#1}\left( #2 \right)}
\newcommand{\iden}{\mathbb{1}}
\renewcommand{\v}[1]{\ensuremath{\boldsymbol #1}}
\newcommand{\norm}[1]{\left\lVert#1\right\rVert}

\theoremstyle{plain}
\newtheorem{theorem}{Theorem}

\newtheorem{corollary}[theorem]{Corollary}
\theoremstyle{definition}
\newtheorem{definition}{Definition}
\theoremstyle{remark}

\newtheorem{example}{Example}

\begin{document}

\title{Structure of the thermodynamic arrow of time in classical and quantum theories} 
\author{Kamil Korzekwa}
\affiliation{Department of Physics, Imperial College London, London SW7 2AZ, United Kingdom}
\affiliation{Centre for Engineered Quantum Systems, School of Physics, The University of Sydney, Sydney, NSW 2006, Australia}
\begin{abstract}
	In this work we analyse the structure of the thermodynamic arrow of time, defined by transformations that leave the thermal equilibrium state unchanged, in classical (incoherent) and quantum (coherent) regimes. We note that in the infinite-temperature limit the thermodynamic ordering of states in both regimes exhibits a lattice structure. This means that when energy does not matter and the only thermodynamic resource is given by information, the thermodynamic arrow of time has a very specific structure. Namely, for any two states at present there exists a unique state in the past consistent with them and with all possible joint pasts. Similarly, there also exists a unique state in the future consistent with those states and with all possible joint futures. We also show that the lattice structure in the classical regime is broken at finite temperatures, i.e., when energy is a relevant thermodynamic resource. Surprisingly, however, we prove that in the simplest quantum scenario of a two-dimensional system, this structure is preserved at finite temperatures. We provide the physical interpretation of these results by introducing and analysing the \emph{history erasure process}, and point out that quantum coherence may be a necessary resource for the existence of an optimal erasure process.

	
\end{abstract}


\maketitle

\section{Introduction}

When considering thermodynamics one inevitably thinks of concepts such as heat flows, thermal machines and work, which seem to be far removed from the ideas of quantum information theory. However, on a more abstract level, thermodynamics can be seen as a field studying the accessibility and inaccessibility of one physical state from another~\cite{caratheodory1909untersuchungen,giles1964mathematical}. The first and second laws of thermodynamics are fundamental constraints on state transformations, forcing thermodynamic processes to conserve the overall energy and forbidding free conversion of heat into work. The mathematical framework developed within quantum information to study the influence of such constraints on the evolution of systems is known under the collective name of \emph{resource theories}. Every resource theory is defined by a subset of quantum states and a subset of quantum operations that are considered free, e.g., in the resource theory of entanglement these are represented by separable states and local operations and classical communication (LOCC), respectively~\cite{horodecki2009quantum}. Every state that is not free is a resource and free transformations, the only ones allowed within the theory, must map the set of free states onto itself. This way no resources can be freely created and one can investigate how they can be exploited and manipulated under free transformations. 

In particular, a set of free thermodynamic operations encodes the structure of the \emph{thermodynamic arrow of time}: it tells us which states can be reached from a given state (and which states can evolve into it) in accordance with the laws of thermodynamics. In other words, whenever a state $\rho$ can freely, i.e., without using any extra thermodynamic resources, evolve to a state $\sigma$, then $\rho$ precedes $\sigma$ in a thermodynamic order. In classical equilibrium thermodynamics in the presence of a single heat bath at fixed temperature such an ordering of states is particularly simple. A free transformation (one that does not require investing work) from one thermal equilibrium state $A$ to another $B$ is possible if and only if $B$ has lower free energy than $A$. The ordering between equilibrium states is thus fully specified by one function and we deal with a total order: either $A$ can be freely transformed to $B$, or $B$ can be freely transformed to $A$. However, this picture becomes more complicated when one considers transformations between different nonequilibrium states. The thermodynamic ordering between incoherent states\footnote{Within this work we refer to states that are incoherent in the energy eigenbasis as simply ``incoherent'' or ``classical'' states.} that arises within the resource theory of thermodynamics, when free transformations are given by catalytic thermal operations, was studied in Ref.~\cite{brandao2013second}. It was proven there that a transformation between incoherent states $\rho$ and $\sigma$ is possible if and only if a whole family of functions, the $\alpha$-R{\'e}nyi divergences between a given state and a thermal equilibrium state, is lower for the final state. Hence, we deal with a partial order, as not every two states are comparable, i.e., it may happen that neither $\rho$ can be transformed into $\sigma$, nor the other way round. Moreover, the ordering arising within the resource theory of thermodynamics between general states with coherence is not a simple generalisation of the ordering between incoherent states. Due to time-translation covariance of thermal operations, a whole new structure emerges where coherence is an independent resource~\cite{lostaglio2015description}. Despite partial results~\cite{lostaglio2015quantum,cwiklinski2015limitations,narasimhachar2015low}, this ordering is still not fully understood.

In this paper, we investigate the thermodynamic ordering that emerges when the set of free transformations is defined by the largest set of quantum operations that do not allow one to construct a perpetuum mobile. These are given by transformations that leave the thermal equilibrium state unchanged, the so-called \emph{Gibbs-preserving}~(GP) operations, and form a superset of thermal operations. Instead of studying the ordering induced by these transformations between particular states, here our focus is on the global properties of the thermodynamic arrow of time. More precisely, a partial order is a very general structure, studied within the field of mathematics known as order theory, with three defining properties: reflexivity, transitivity and antisymmetry. Being such a broad and general concept, it seems natural to ask whether the thermodynamic ordering has a more rigid and specific structure. Inspired by order-theoretic studies we will focus on a special kind of partial order known as a \emph{lattice} and interpret it from a thermodynamic perspective. We will show that in the infinite-temperature limit the thermodynamic arrow of time actually reflects the structure of a lattice, but that this structure is lost in the classical regime at finite temperatures. Surprisingly, however, we will also prove that in the simplest quantum scenario of a two-dimensional system, the lattice structure is preserved at finite temperatures. This suggests that coherence can play a role in providing structure to the thermodynamic arrow of time. Finally, we will introduce and analyse the \emph{history erasure process} in order to provide a physical interpretation of these results and to highlight the differences between the infinite- and finite-temperature cases, as well as between classical and quantum scenarios.

\section{Setting the scene}

\subsection{Thermodynamic framework}

We will investigate the thermodynamics of finite-dimensional systems of dimension $d$ in the presence of a single heat bath at inverse temperature $\beta=1/k_B T$, where $k_B$ denotes the Boltzmann constant. The space of quantum states is then given by the set of density operators $\Sc_d$, i.e., the set of positive-semidefinite operators $\rho$ with unit trace that act on a $d$-dimensional Hilbert space. Among all the bases of $\Sc_d$ the energy eigenbasis $\{\ketbra{E_i}{E_j}\}$, defined by the eigenstates of the system Hamiltonian $H=\sum_i E_i \ketbra{E_i}{E_i}$ with $E_i\leq E_{i+1}$, is distinguished by the evolution of the system. States diagonal in this basis evolve trivially in time under the free evolution of the system, i.e., they are time-translation invariant. Such incoherent states are usually referred to as \emph{classical states} because their energy is well defined up to observer's classical lack of knowledge and they do not exhibit quantum features of superposition. This is in strict contrast to generic quantum states with coherence between energy eigenstates, when even the possession of complete knowledge about a state may leave one with uncertainty about its energy. A classical state can be equivalently described by a probability vector $\v{p}$ corresponding to the diagonal of a density matrix written in the energy eigenbasis: $p_i=\matrixel{E_i}{\rho}{E_i}$. Therefore, the space of classical states can be identified with the set of $d$-dimensional probability distributions $\P_d$.

The most general evolution of a quantum state $\rho$ of a system that is initially uncorrelated with its environment is given by a \emph{completely positive trace-preserving} (CPTP) map~$\E$, also known as a quantum channel. Similarly, the most general evolution between two classical states $\v{p}$ and $\v{q}$ is described by a stochastic matrix~$\Lambda$. However, due to thermodynamic constraints encoded by the laws of thermodynamics, not all state transformations are allowed, e.g., a system cannot be taken out of thermal equilibrium for free. In order to define a resource theory that captures these thermodynamic limitations we need to identify free states and free operations of the theory. By definition, a state of a system that is in equilibrium with a thermal bath at inverse temperature $\beta$ is a free state. Therefore, for a system described by a Hamiltonian $H$, the only free state is the \emph{thermal Gibbs state}, 
\begin{equation}
\gamma=\frac{e^{-\beta H}}{Z},\quad Z=\tr{}{e^{-\beta H}}.
\end{equation}
Note that, being diagonal in the energy eigenbasis, $\gamma$ is a classical state and thus can be represented by a probability vector $\v{\gamma}$. To define free thermodynamic operations we use the fact that they must map the set of free states into itself, so that they must preserve~$\gamma$. The property of having a fixed point $\gamma$ is thus a minimal requirement on any set of free thermodynamic operations. It enforces all states to evolve towards the thermal equilibrium state $\gamma$ and ensures that one cannot bring a system out of equilibrium at no work cost. Indeed, if this was not the case, we could equilibrate it back and extract work, thus constructing a perpetuum mobile and breaking the second law of thermodynamics. We conclude that the largest set of operations that are consistent with the thermodynamic arrow of time must preserve the Gibbs state and is defined as follows:
\begin{definition}[Gibbs-preserving maps]\label{def:GP_maps}
	We say that a CPTP map $\E$ is \textit{Gibbs-preserving}, or GP for short, if the thermal Gibbs state $\gamma$ is its fixed point: $\E(\gamma)=\gamma$. Similarly, in the classical case, we say that a stochastic matrix $\Lambda$ is GP if the thermal equilibrium probability distribution $\v{\gamma}$ is its fixed point: $\Lambda\v{\gamma}=\v{\gamma}$.
\end{definition}

\begin{figure}[t!]
	\includegraphics[width=0.72\columnwidth]{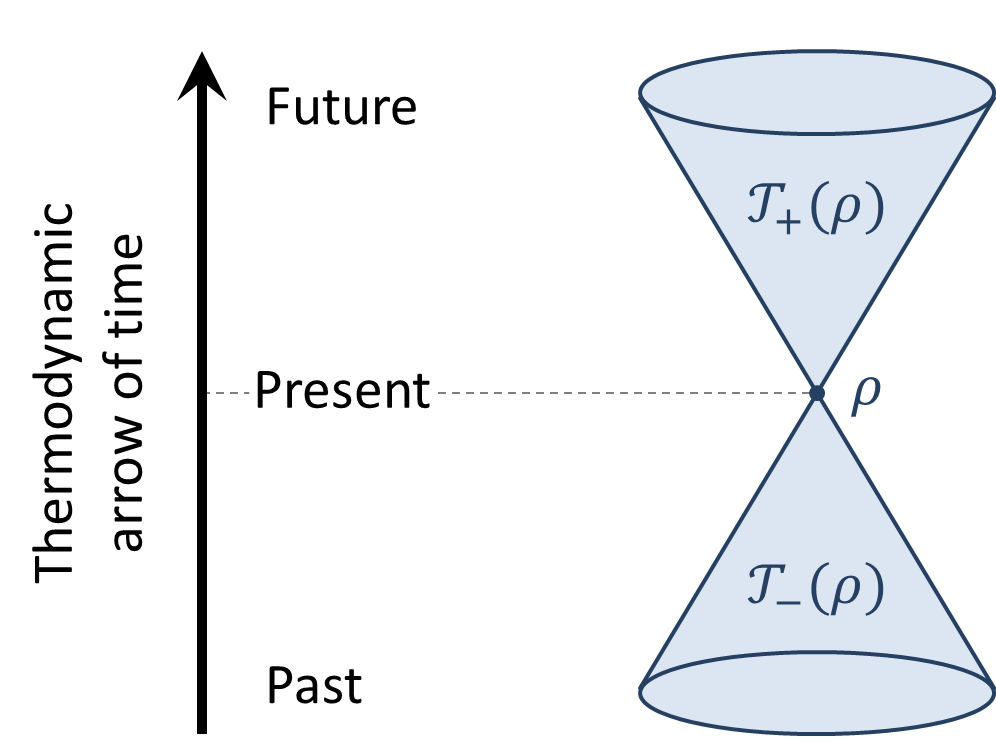}
	\caption{\label{fig:thermal_cones} \emph{Thermal cones.} The reachability of one state from another via a GP map introduces the ordering of states along the thermodynamic arrow of time. States that can be reached from a given state $\rho$ form its future thermal cone~$\T_+(\rho)$, whereas states that can be transformed into $\rho$ form its past thermal cone~$\T_-(\rho)$.}
\end{figure}

In order to investigate the thermodynamic ordering of states encoded by Gibbs-preserving maps we will use the notion of \emph{thermal cones}, defined as follows.
\begin{definition}[Thermal cones]
The set of states $\T_+(\rho)$ that a quantum state $\rho$ can be mapped to via GP quantum channels is called the \textit{future thermal cone} of $\rho$. The set of states $\T_-(\rho)$ that can be mapped to $\rho$ via GP quantum channels is called the \textit{past thermal cone} of $\rho$. The future and past thermal cones of a classical state $\v{p}$ are defined in an analogous way, simply by replacing GP quantum channels with GP stochastic matrices.
\end{definition}
\noindent Thermal cones induce ordering along the thermodynamic arrow of time within the state space (see Fig.~\ref{fig:thermal_cones}). In order to rigorously investigate the structure of this ordering, in the next section we will introduce the necessary concepts from the field of order theory. As the following considerations can be applied to both the classical and quantum cases, for the clarity of discussion we will just state them for the more general quantum case. To obtain classical statements one simply needs to replace a density matrix $\rho\in\Sc_d$ with a probability vector $\v{p}\in\P_d$ and a GP quantum channel $\E$ with a GP stochastic matrix $\Lambda$.

\subsection{Order theoretic approach}

\subsubsection{Partial order}
\label{sec:partial_order}

Since the identity operation is a Gibbs-preserving CPTP map and a composition of two GP operations is also GP, we have that $\rho\in\T_\pm(\rho)$ and
\begin{equation*}
\left(\rho\in\T_\pm(\sigma)\mathrm{~and~}\sigma\in\T_\pm(\tau)\right)\Rightarrow\rho\in\T_\pm(\tau).
\end{equation*}
Hence, the relation of one state belonging to a thermal cone of another state is reflexive and transitive. Such binary relations are known as \textit{preorders} and are usually denoted by~$\succsim$. The set of all past thermal cones induces a preorder~$\succsim_-$ ``oriented along'' the thermodynamic arrow of time, i.e., $\rho$ precedes $\sigma$, $\rho\succsim_-\sigma$, if $\rho\in\T_-(\sigma)$. On the other hand, the set of all future thermal cones induces a preorder~$\succsim_+$ ``oriented against'' the thermodynamic arrow of time, i.e., $\rho$ precedes $\sigma$, $\rho\succsim_+\sigma$, if $\rho\in\T_+(\sigma)$. As these two preorders are dual, meaning that $\rho\succsim_+\sigma$ is equivalent to $\sigma\succsim_-\rho$, without loss of generality we will focus only on the preorder~$\succsim_-$ oriented along the thermodynamic arrow of time and simply denote it by~$\succsim$. In other words, instead of writing $\rho\in\T_-(\sigma)$ we can write $\rho\succsim\sigma$, meaning that $\rho$ precedes $\sigma$ in the thermodynamic preorder.

For two states $\rho$ and $\sigma$ it may happen that $\rho\succsim\sigma$ and $\sigma\succsim\rho$. We then say that $\rho$ and $\sigma$ are \emph{reversibly interconvertible} under GP operations and denote it by $\rho\sim\sigma$. Since the relation~$\sim$ is reflexive, transitive and symmetric, it is an equivalence relation and thus the set of states that are reversibly interconvertible forms a \emph{thermodynamic equivalence class}.
\begin{figure}[t!]
	\includegraphics[width=0.85\columnwidth]{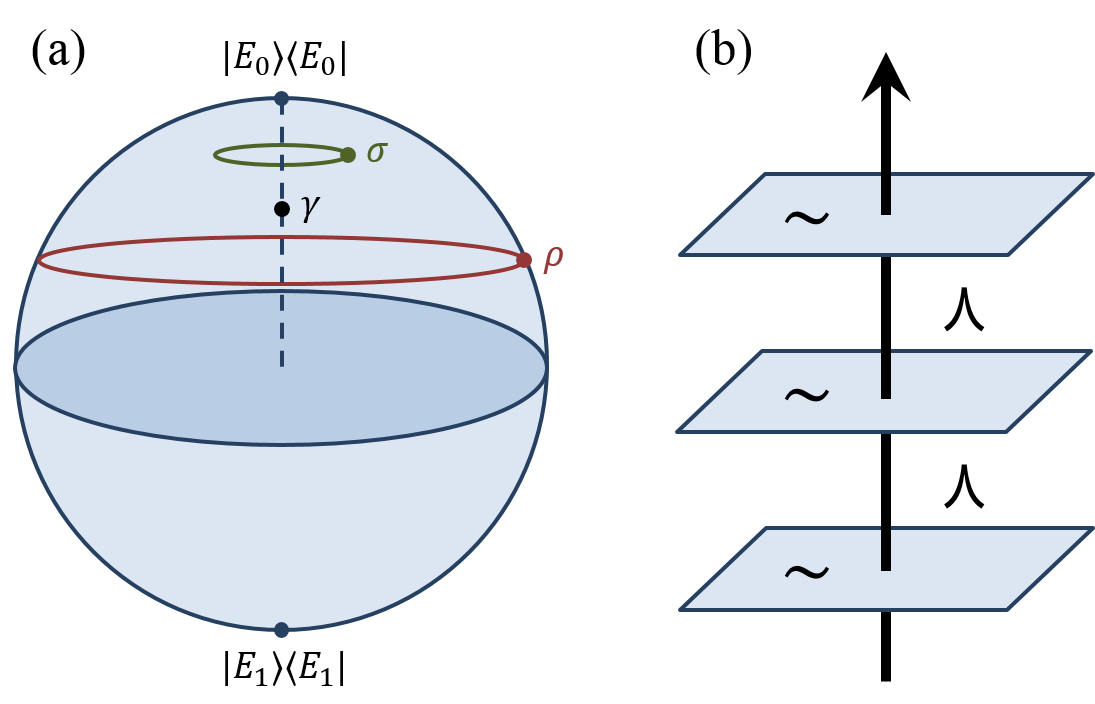}
	\caption[Decomposition of preorder into partial order between equivalence classes]{\label{fig:preorder_decomposition}\emph{Decomposition of preorder into partial order between equivalence classes.} (a) States on the unitary orbit generated by $U(t)=e^{-iHt}$ are all reversibly interconvertible under GP transformations, as $U(t)$ is GP. Hence, if a state $\rho$ can thermodynamically evolve to $\sigma$, then any state $U(t)\rho U^\dagger(t)$ can evolve to any state $U(t')\sigma U^\dagger(t')$. Thus, without loss of generality one can only study the ordering between single representatives of each orbit. (b) States that are reversibly interconvertible by GP operations can be thought of as living in a plane ``perpendicular'' to the thermodynamic arrow of time, i.e., members of a thermodynamic equivalence class are at the same stage of the evolution towards equilibrium. The thermodynamic ordering of states is then given by the partial order between planes.}
\end{figure}
Using thermodynamic equivalence classes we can simplify thermodynamic preorders in the following way. Instead of considering the ordering between all states, we can restrict our study to the ordering between single representatives of each thermodynamic equivalence class. This results in promoting the thermodynamic preorder $\succsim$ into a partial order $\succ$, which has an additional property of being antisymmetric, i.e., if $\rho\succ\sigma$ and $\sigma\succ\rho$ then $\rho=\sigma$. Let us explain this construction with the use of the following example:

\begin{example}[Thermodynamic equivalence class]
	Consider a qubit system described by a Hamiltonian \mbox{$H=\sigma_z$} prepared in a general state with Bloch vector \mbox{$\v{r}=(r\cos\phi,r\sin\phi,z)$}. Since a unitary \mbox{$U(t)=e^{-iHt}$}, mapping $\v{r}$ to \mbox{$\v{r}'=(r\cos\phi',r\sin\phi',z)$}, is a reversible GP map, states with fixed $r$ and $z$ belong to the same thermodynamic equivalence class. Hence, we can focus only on one representative of each class parametrised by $r$ and $z$, e.g., with $\phi=0$ (see Fig.~\ref{fig:preorder_decomposition}).
\end{example} 
\noindent In fact, we have just used a standard procedure that allows one to decompose any preorder~$\succsim$ on a generic set $S$ into a partial order~$\succ$ between equivalence classes (subsets of~$S$ with elements connected via an equivalence relation~$\sim$). Let us summarise this section by stating a formal definition of thermodynamic ordering:

\begin{definition}[Thermodynamic ordering]\label{def:thermo_order}
	Consider a preorder~$\succsim$ on the set~$\Sc_d$ of quantum states defined by a relation of belonging to the past thermal cone, i.e., $\rho\succsim\sigma$ if and only if $\rho\in\T_-(\sigma)$. Identify each set of states that are reversibly interconvertible via GP maps, i.e., $\rho\succsim\sigma$ and $\sigma\succsim\rho$, with a thermodynamic equivalence class~$\sim$. Thermodynamic ordering is a partial order~$\succ$ between those equivalence classes, i.e., a partial order on the quotient set~$\Sc_d/\sim$.
\end{definition}

\subsubsection{Lattice}

\begin{figure}[t!]
	\begin{centering}
		\includegraphics[width=\columnwidth]{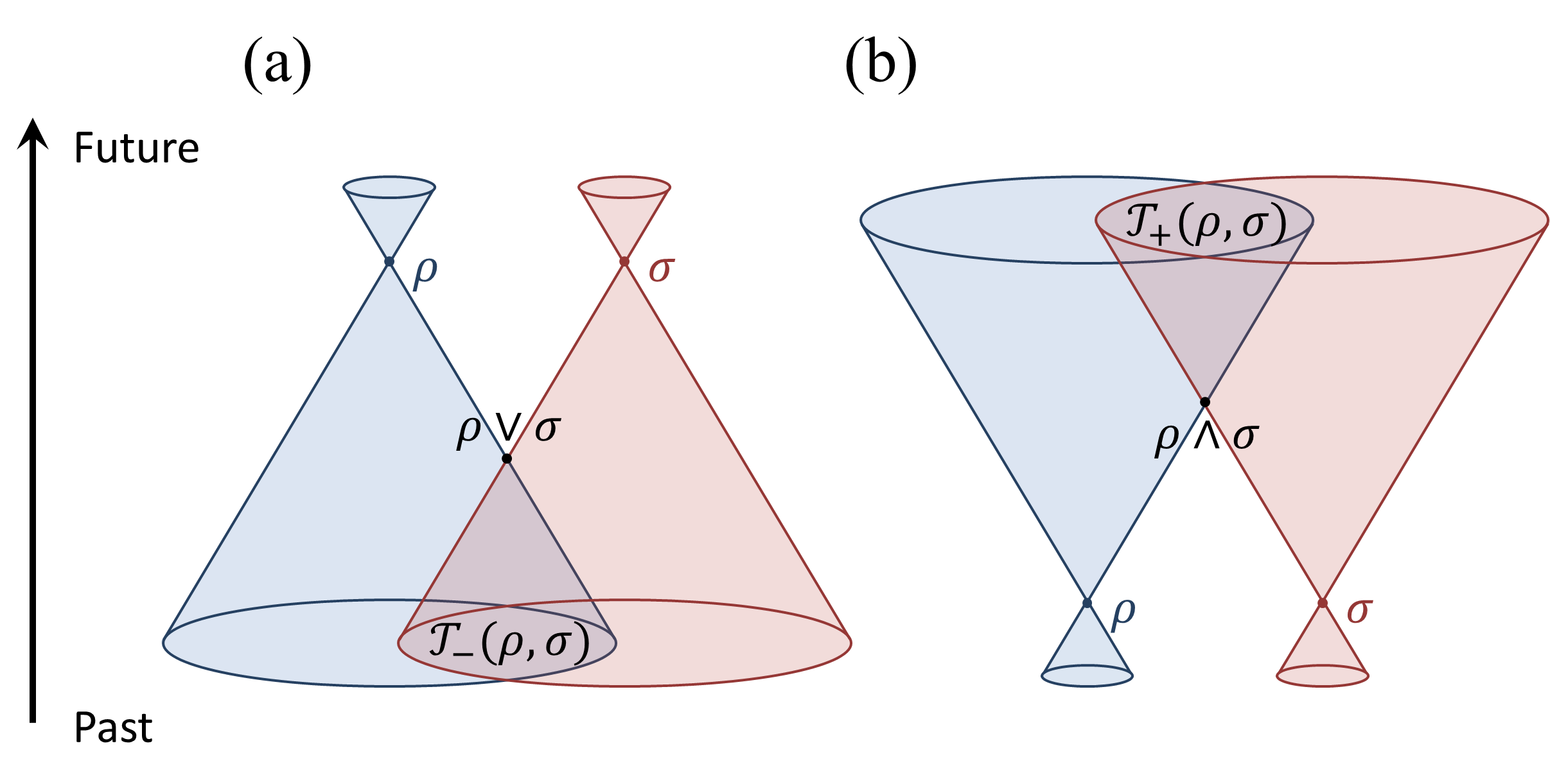}
		\caption[Visualising the join and meet]{\emph{Visualising the join and meet.} (a) The intersection of past thermal cones of $\rho$ and $\sigma$, denoted by $\T_-(\rho,\sigma)$, is a set of states that can thermodynamically evolve to both $\rho$ and $\sigma$. The join $\rho\vee\sigma$ is the unique state belonging to $\T_-(\rho,\sigma)$ that can be thermodynamically reached from all states in $\T_-(\rho,\sigma)$. (b) The intersection of future thermal cones of $\rho$ and $\sigma$, denoted by $\T_+(\rho,\sigma)$, is a set of states that can be thermodynamically reached from both $\rho$ and $\sigma$. The meet $\rho\wedge\sigma$ is the unique state belonging to $\T_+(\rho,\sigma)$ that can thermodynamically evolve to all states in $\T_+(\rho,\sigma)$.\label{fig:join_meet}}
	\end{centering}
\end{figure}

We will now provide a definition and interpret a special kind of partial order: a lattice. However, before we can do it we first need to introduce a few more notions, which are illustrated in Fig.~\ref{fig:join_meet}. For any two states $\rho$ and $\sigma$ let us introduce a set of states \mbox{$\T_-(\rho,\sigma)=\T_-(\rho)\cap\T_-(\sigma)$}, i.e., the set of all states whose future thermal cones contain both $\rho$ and $\sigma$. The thermodynamic interpretation of $\T_-(\rho,\sigma)$ is that of a set of states in the past that are allowed by the thermodynamic arrow of time to evolve both into $\rho$ and $\sigma$ at present. Similarly, let us introduce a set of states \mbox{$\T_+(\rho,\sigma)=\T_+(\rho)\cap\T_+(\sigma)$}, i.e., the set of all states whose past thermal cones contain both $\rho$ and $\sigma$. Thermodynamically $\T_+(\rho,\sigma)$ is the set of states in the future that are allowed by the thermodynamic arrow of time to be reached from both $\rho$ and $\sigma$ at present.

Now, if there exists $\tau_-\in\T_-(\rho,\sigma)$ such that for all $\tau\in\T_-(\rho,\sigma)$ we have $\tau\succ\tau_-$ then $\tau_-$ is called the \textit{join} of $\rho$ and $\sigma$ and is usually denoted by $\rho\vee\sigma$. The notation is justified by the fact that $\T_+(\tau_-)$ is the smallest thermal cone that contains \mbox{$\T_+(\rho)\cup\T_+(\sigma)$}. Thermodynamically we can interpret the join of $\rho$ and $\sigma$ as the unique state\footnote{It is important to remember that, precisely speaking, the thermodynamic ordering is a partial order between thermodynamic equivalence classes, as explained in Definition~\ref{def:thermo_order}, and not between all states. Hence, whenever we refer to a ``unique state'', we actually mean a state that is unique up to equivalence relation. In other words, we identify all states that are reversibly interconvertible under GP operations with one representative state that represents all states within this particular equivalence class.} in the past that is consistent both with $\rho$ and $\sigma$ at present, as well as with all possible joint pasts of $\rho$ and $\sigma$. The join can also be seen as the extremal moment in the past evolution, at which the system has to ``decide'' whether to evolve into $\rho$ or $\sigma$. 

Similarly, if there exists $\tau_+\in\T_+(\rho,\sigma)$ such that for all $\tau\in\T_+(\rho,\sigma)$ we have $\tau_+\succ\tau$, then $\tau_+$ is called the \textit{meet} of $\rho$ and $\sigma$ and is usually denoted by $\rho\wedge\sigma$. Again, the notation is justified by the fact that \mbox{$\T_+(\tau_+)$} is the biggest thermal cone that is contained in \mbox{$\T_+(\rho)\cap\T_+(\sigma)$}. Thermodynamically the meet of $\rho$ and $\sigma$ is the unique state in the future that is consistent both with $\rho$ and $\sigma$ at present, as well as with all possible joint futures of $\rho$ and $\sigma$. The meet can also be seen as the extremal moment in the future evolution, after which the system ``forgets'' whether it evolved from $\rho$ or $\sigma$, as its state is consistent with both pasts.

\begin{definition}[Thermodynamic lattice]\label{def:thermo_lattice}
	The thermodynamically ordered set of quantum states $(\Sc_d,\succ)$ forms a thermodynamic lattice if for every pair of states $\rho,\sigma\in\Sc_d$ (defined up to thermodynamic equivalence relation) there exists a join and meet. 
\end{definition}

\begin{figure}[t!]
	\begin{centering}
		\includegraphics[width=\columnwidth]{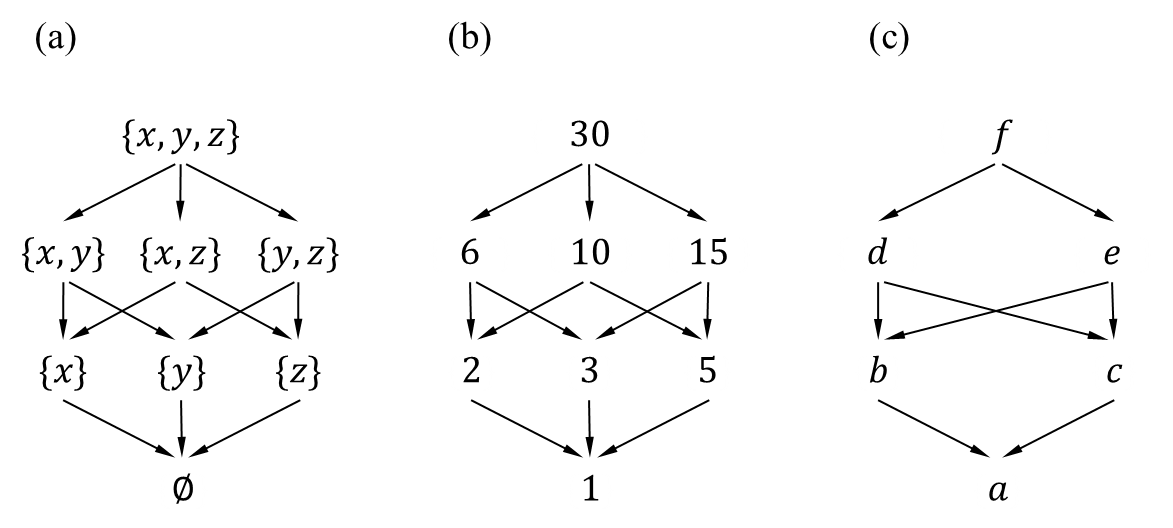}
		\caption[Examples of lattice and non-lattice partial orders]{\emph{Examples of lattice and non-lattice partial orders.} Partially ordered sets can be represented by their Hasse diagrams: an arrow from $a$ to $b$ denotes $a\succ b$. (a) The power set of any set $A$ forms a lattice under the partial order induced by subset inclusion. The join and meet are given by set union and intersection, respectively. Here, we choose a three-element set $A=\{x,y,z\}$. (b) Natural numbers partially ordered by divisibility, i.e., $a\succ b$ if and only if $b$ is a divisor of $a$, form a lattice. The join and meet are given by the operations of taking the least common multiple and greatest common divisor, respectively. Here we present a sublattice of divisors of $30$. (c) A set $\{a,b,c,d,e,f\}$ with partial ordered defined by the presented Hasse diagram does not form a lattice. Although $b$ and $c$ have common upper bounds $d$, $e$, and $f$, neither of them is a join, i.e., the least upper bound. \label{fig:lattice_examples}}
	\end{centering}
\end{figure}

The existence of a thermodynamic lattice would not only bring a new understanding of the thermodynamic arrow of time (with a unique consistent future and past for each subset of states), but could also allow us to use new algebraic tools to study thermodynamics. Namely, if $(\Sc_d,\succ)$ forms a thermodynamic lattice, then it can be fully described as an algebraic structure $(\Sc_d,\vee,\wedge)$ consisting of a set of quantum states $\Sc_d$ and two binary operations $\vee$ and $\wedge$ (i.e., functions $\Sc_d\times\Sc_d\rightarrow\Sc_d$) satisfying the following axioms for all $\rho,\sigma,\tau\in\Sc_d$:
\begin{subequations}
	\begin{eqnarray}
	\rho\vee\sigma&=&\sigma\vee\rho,\\
	\rho\vee(\sigma\vee\tau)&=&(\rho\vee\sigma)\vee\tau\\
	\rho\vee(\rho\wedge\sigma)&=&\rho,
	\end{eqnarray}
\end{subequations}
and another three obtained from the above by exchanging $\vee$ with $\wedge$. In Fig.~\ref{fig:lattice_examples} we present examples of partial orders that do and do not form a lattice.

\section{Ordering of classical states}

We are now ready to analyse the structure of the thermodynamic ordering of states. We will start with classical states described by probability distributions \mbox{$\v{p}\in\P_d$}, considering separately the infinite-temperature limit, \mbox{$\beta\rightarrow 0$}, and the case of finite temperatures. The former can also be thought of as an information-theoretic limit, because with energy states all being degenerate, negentropy (or information) is the only thermodynamic resource. In this case, as we will show, the thermodynamic arrow of time exhibits the structure of a lattice (known as the information lattice~\cite{cicalese2002supermodularity}). However, we will also prove that as soon as different energy states become distinct, the lattice structure is broken within the classical theory.

\subsection{Infinite temperature and a lattice structure}
\label{sec:infinite_T_lattice}

In the infinite-temperature limit the Gibbs state is described by a uniformly distributed probability vector $\v{\eta}:=(1/d,1/d,\dots,1/d)$. Hence, thermodynamically allowed GP~transformations are given by bistochastic maps such that $\Lambda \v{\eta}=\v{\eta}$. The existence of a bistochastic map $\Lambda$ satisfying \mbox{$\Lambda\v{p}=\v{q}$} is equivalent to $\v{p}\succ\v{q}$~\cite{bhatia1997matrix}, where~$\succ$ denotes majorisation relation defined as follows. Denote by $\v{p}^{\downarrow}$ a probability distribution $\v{p}$ with entries rearranged in non-increasing order. Then, $\v{p}$ is said to majorise $\v{q}$ if and only if
\begin{equation}
\sum_{i=1}^k p_i^{\downarrow}\geq\sum_{i=1}^k q_i^{\downarrow},
\end{equation}
for all $k\in\{1\dots d\}$. Within the space of $d$-dimensional probability distributions~$\P_d$, majorisation forms a preorder, not a partial order, because for two probability distributions $\v{p}$ and $\v{q}$ that are connected by a permutation, $\v{q}=\Pi\v{p}$, we have $\v{p}\succ\v{q}$ and $\v{q}\succ\v{p}$, but $\v{p}\neq\v{q}$. However, as discussed before, we can identify all probability distributions that are connected by some permutation (which is a reversible bistochastic map) with an equivalence class, and then focus on the partial order between those equivalence classes.

\begin{figure}[t!]
	\begin{centering}
		\includegraphics[width=\columnwidth]{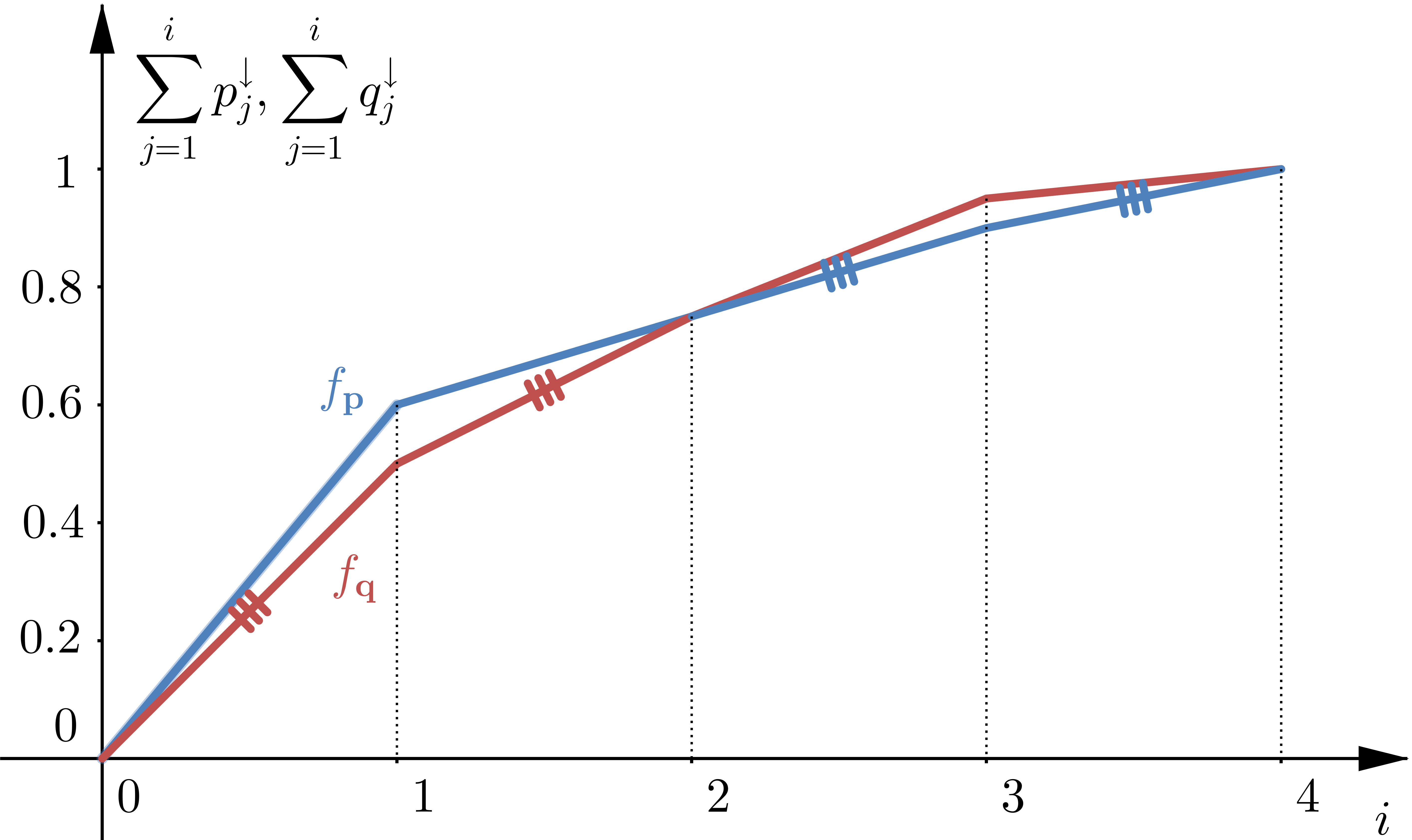}
		\caption[Meet of a majorisation lattice]{\label{fig:meet_majo}\emph{Meet of a majorisation lattice.} Majorisation curves $f_{\v{p}}$ and $f_{\v{q}}$ (solid lines) for $\v{p}=(0.6,0.15,0.15,0.1)$ and $\v{q}=(0.5,0.25,0.2,0.05)$, together with their meet \mbox{$\v{p}\wedge\v{q}=(0.5,0.25,0.15,0.1)$} (given by the line consisting of triple-crossed segments).}
	\end{centering}
\end{figure}

It is known that majorisation partial order forms a lattice: for any two probability distributions, $\v{p}$ and $\v{q}$, there exists meet $\v{p}\wedge\v{q}$ and join $\v{p}\vee\v{q}$~\cite{cicalese2002supermodularity}. More precisely, the meet is defined as a probability vector $\v{l}$ with components given by
\begin{equation}
\label{eq:majo_meet}
l_i=\min\left\{\sum_{j=1}^i p_j^\downarrow,\sum_{j=1}^i q_j^\downarrow\right\}-\min\left\{\sum_{j=1}^{i-1} p_j^\downarrow,\sum_{j=1}^{i-1} q_j^\downarrow\right\}.
\end{equation}
To explain why this is the case, let us use majorisation curves, $f_{\v{p}}$ and $f_{\v{q}}$, that consist of linear segments connecting points $(i,\sum_{j=1}^i p_j^{\downarrow})$ for $i\in\{0,\dots,d\}$, and analogously for $\v{q}$. Now, the condition $\v{p}\succ\v{q}$ is equivalent to $f_{\v{p}}\geq f_{\v{q}}$ everywhere. Hence, for the meet $\v{l}=\v{p}\wedge\v{q}$, its majorisation curve $f_{\v{l}}$ must be the ``maximal'' curve lying below both $f_{\v{p}}$ and $f_{\v{q}}$. The expression in Eq.~\eqref{eq:majo_meet} ensures that, as the resulting majorisation curve $f_{\v{l}}$ is equal to $\min\{f_{\v{p}},f_{\v{q}}\}$ at each point. We illustrate this for exemplary probability vectors in Fig.~\ref{fig:meet_majo}. 

\begin{figure}[t!]
	\begin{centering}
		\includegraphics[width=\columnwidth]{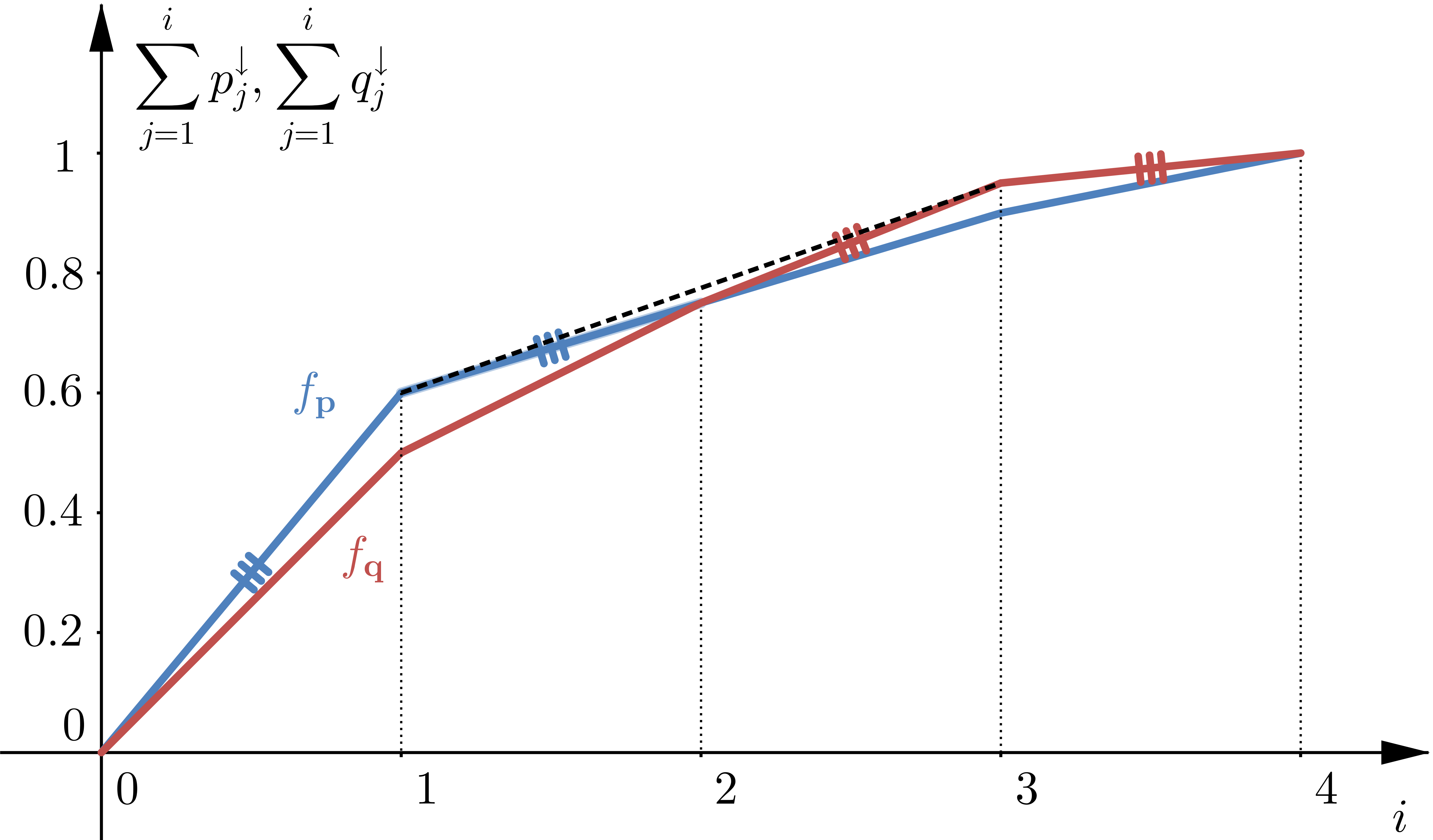}
		\caption[Join of a majorisation lattice]{\label{fig:join_majo}\emph{Join of a majorisation lattice.} Majorisation curves $f_{\v{p}}$ and $f_{\v{q}}$ (solid lines) for \mbox{$\v{p}=(0.6,0.15,0.15,0.1)$} and \mbox{$\v{q}=(0.5,0.25,0.2,0.05)$}, together with the curve corresponding to \mbox{$\v{g}^{(0)}=(0.6,0.15,0.2,0.05)$} from Eq.~\eqref{eq:majo_join1} (given by the line consisting of triple-crossed segments). This curve is not concave, since $\v{g}^{(0)}$ is not arranged in a nonincreasing order. Hence, in order to obtain the join $\v{p}\vee\v{q}$, one needs to ``smooth'' the curve between the points $i=1$ and $i=3$ (dashed black line). This leads to the join given by \mbox{$\v{p}\vee\v{q}=(0.6,0.175,0.175,0.05)$}.}
	\end{centering}
\end{figure}

The join $\v{p}\vee\v{q}$ can be constructed with the use of the following algorithm~\cite{cicalese2002supermodularity}. First, define a probability vector $\v{g}^{(0)}$ with components given by
\begin{equation}\small
\label{eq:majo_join1}
g_i^{(0)}=\max\left\{\sum_{j=1}^i p_j^\downarrow,\sum_{j=1}^i q_j^\downarrow\right\}-\max\left\{\sum_{j=1}^{i-1} p_j^\downarrow,\sum_{j=1}^{i-1} q_j^\downarrow\right\}.
\end{equation}\normalsize
Now, the iterative application of the following transformation on $\v{g}^{(k)}$ results in $\v{p}\vee\v{q}$ in no more than $d-1$ steps. Start with $k=0$. Denote by $n\geq 2$ the smallest integer such that $g_n^{(k)}>g_{n-1}^{(k)}$ and by $m\leq n-1$ the greatest integer such that 
\begin{equation}
\label{eq:majo_join2}
g_{m-1}^{(k)}\geq\frac{\sum_{i=m}^n g_i^{(k)}}{n-m+1}=:a_k.
\end{equation}
Define $\v{g}^{(k+1)}$ by setting its components $g_i^{(k+1)}=a_k$ for $i\in\{m\dots n\}$ and $g_i^{(k+1)}=g_i^{(k)}$ otherwise. Repeat until for some $k'$ the vector $\v{g}^{(k')}$ has components ordered in a nonincreasing order. The join $\v{p}\vee\v{q}$ is then given by $\v{g}^{(k')}$. The procedure just described starts similarly to the one used to define meet: we introduce a state $\v{g}^{(0)}$, whose majorisation curve is the ``minimal'' curve lying above both $f_{\v{p}}$ and $f_{\v{q}}$, i.e., $f_{\v{g}^{(0)}}$ is equal to $\max\{f_{\v{p}},f_{\v{q}}\}$ at each point. The problem is that the resulting curve may not be concave, and since the majorisation curve is constructed from the components arranged in a nonincreasing order, each such curve must be concave. What the described algorithm does to overcome this problem, is to identify points at which the curve breaks concavity, and ``smooth'' it over sufficient number of points to guarantee concavity. We illustrate this for exemplary probability vectors in Fig.~\ref{fig:join_majo}.

\subsection{Finite temperatures and a broken lattice structure}
\label{sec:finite_temp}

At finite temperatures a classical state $\v{p}$ can be mapped via a GP stochastic map $\Lambda$ into $\v{q}$ if and only if $\v{p}$ thermo-majorises $\v{q}$~\cite{ruch1978mixing,horodecki2013fundamental}. The thermo-majorisation partial order is usually denoted by $\succ_\beta$ and is defined in the following way. First, we need the notion of \mbox{\emph{$\beta$-ordering}} of probability distributions. Given a thermal Gibbs distribution $\v{\gamma}$, with a fixed inverse temperature $\beta$, introduce a \emph{Gibbs-rescaled} version of $\v{p}$: $\v{p}^\gamma=(p_1/\gamma_1,\dots,p_d/\gamma_d)$. The \mbox{$\beta$-ordering} of $\v{p}$ is defined by a permutation $\pi_{\v{p}}$ that arranges $\v{p}^\gamma$ in a nonincreasing order, i.e.,
	\begin{equation}
	\left(\v{p}^\gamma\right)^\downarrow=\left(p^\gamma_{\pi_{\v{p}}^{-1}(1)},\dots,p^\gamma_{\pi_{\v{p}}^{-1}(d)}\right).
	\end{equation}
	Now, the $\beta$-ordered version of a probability vector $\v{p}$ is given by
	\begin{equation}
	\v{p}^\beta=\left(p_{\pi_{\v{p}}^{-1}(1)},\dots,p_{\pi_{\v{p}}^{-1}(d)}\right).
	\end{equation}	
We say that two probability distributions $\v{p}$ and $\v{q}$ belong to the same $\beta$-ordering if the same permutation matrix rearranges their Gibbs-rescaled versions into a nonincreasing order.	Next, for every probability distribution $\v{p}$ we also define a \emph{thermo-majorisation curve} $f_{\v{p}}$, which is composed of linear segments connecting the point $(0,0)$ and the points
\begin{equation}
\left(\sum_{i=1}^k\gamma^\beta_i,~\sum_{i=1}^k p^\beta_i\right)=\left(\sum_{i=1}^k\gamma_{\pi^{-1}_{\v{p}}(i)},~\sum_{i=1}^k p_{\pi^{-1}_{\v{p}}(i)}\right),
\end{equation}
for $k\in\{1,\dots,d\}$, where $\pi_{\v{p}}$ is a permutation that $\beta$-orders $\v{p}$. Finally, $\v{p}$ thermo-majorises $\v{q}$ if the thermo-majorisation curve $f_{\v{p}}$ is above $f_{\v{q}}$ everywhere, i.e., $f_{\v{p}}(x)\geq f_{\v{q}}(x)$.

We will now first show that in the simplest case of a two-level system the partial order induced by thermo-majorisation on the full state space $\P_2$ does not form a lattice. Then, we will present how to generalise this result for~$d>2$. However, we will also prove that a lattice structure is preserved within subspaces of $\P_d$ containing probability vectors belonging to the same $\beta$-ordering. In other words, for two states $\v{p}$ and $\v{q}$ belonging to the same $\beta$-ordering, there may be many incomparable ``candidate'' states for meet and join; but within the subset consisting only of probability vectors with the same $\beta$-ordering as $\v{p}$ and $\v{q}$, the meet and join will be defined uniquely.

For any given two-dimensional thermal state \mbox{$\v{\gamma}=(\gamma_0,1-\gamma_0)$} with $\gamma_0\neq1/2$, let us choose two states, \mbox{$\v{p}=(p,1-p)$} and \mbox{$\v{q}=(q,1-q)$}, with
\begin{equation}
p=\frac{1+\gamma_0}{2}\geq \gamma_0,\quad q=\frac{2\gamma_0-1}{\gamma_0}\leq \gamma_0.
\end{equation}
We will prove that there does not exist a join for these two states. A generic two-level system is described by $\v{r}=(r,1-r)$. It is easy to verify that for $r\geq \gamma_0$ the only state that thermo-majorises both $\v{p}$ and $\v{q}$ is given by $r=1$. On the other hand, for $r\leq \gamma_0$ we get that $\v{r}$ thermo-majorises both $\v{p}$ and $\v{q}$ if $r\leq \gamma_0/2$. Among these states the ones with \mbox{$r<\gamma_0/2$} thermo-majorise the one with \mbox{$r=\gamma_0/2$}. Hence, we are left only with two candidates for the join of $\v{p}$ and $\v{q}$, namely $(1,0)$ and $(\gamma_0/2,1-\gamma_0/2)$. By direct inspection we find that those states are incomparable under thermo-majorisation partial order, so no join exists and we do not have a lattice structure. In a similar fashion one can prove that there always exists states $\v{p}$ and $\v{q}$ for which no meet exists [for example, by choosing $p=(3+\gamma_0)/4$ and $q=(\gamma_0^2+2\gamma_0-1)/(4\gamma_0)$].

\begin{figure}[t!]
	\begin{centering}
		\includegraphics[width=0.95\columnwidth]{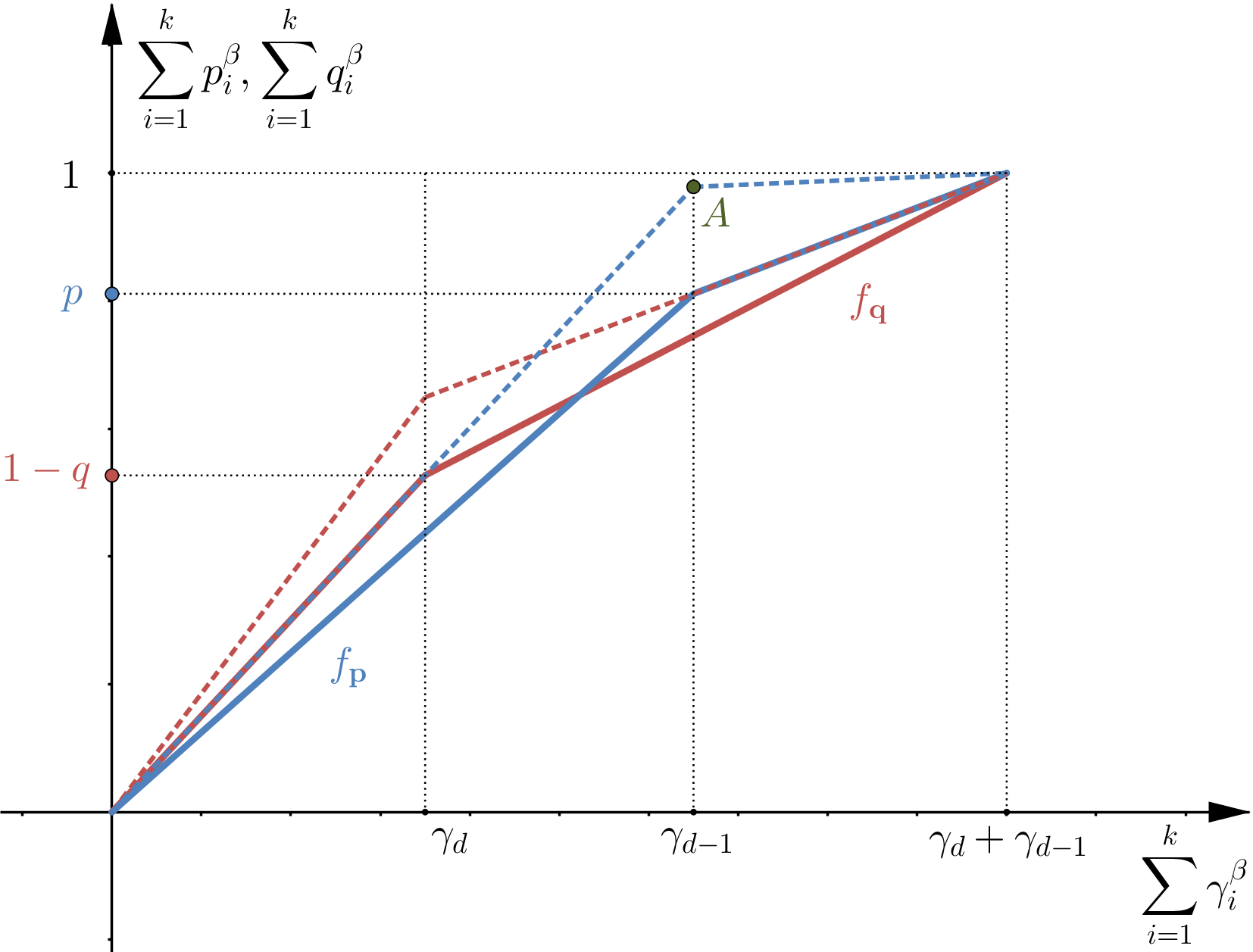}
		\caption[Thermo-majorisation order is not a lattice]{\label{fig:thermomajorisation}\emph{Thermo-majorisation order is not a lattice.} Thermo-majorisation curves $f_{\v{p}}$ and $f_{\v{q}}$ (solid lines), together with the candidates for the join $\v{p}\vee\v{q}$, i.e., optimal curves (plotted with dashed lines) thermo-majorising both $\v{p}$ and $\v{q}$. Satisfying the inequalities given by Eq.~\eqref{eq:ineq} guarantees that: 1.~$f_{\v{q}}>f_{\v{p}}$ at $\gamma_d$; 2.~$f_{\v{q}}<f_{\v{p}}$ at $\gamma_{d-1}$; 3.~Point $A$ lies below 1, which results in the existence of two incomparable candidates for $\v{p}\vee\v{q}$.}
	\end{centering}
\end{figure}

Let us now consider two $d$-dimensional probability distributions \mbox{$\v{p}=(0,\dots,0,p,1-p)$} and \mbox{$\v{q}=(0,\dots,0,q,1-q)$}, i.e., classical states with only the two highest energy levels $E_{d-1}$ and $E_{d}$ occupied. For simplicity, let us assume that these two energy levels are non-degenerate. This simplifying assumption allows us to map a $d$-dimensional problem to a two-dimensional one, showing that thermo-majorisation does not have the structure of a lattice for $d>2$. Let us now choose $p\geq\frac{\gamma_{d-1}}{\gamma_{d}+\gamma_{d-1}}$ and $q\leq\frac{\gamma_{d-1}}{\gamma_{d}+\gamma_{d-1}}$. This way the corresponding thermo-majorisation curves $f_{\v{p}}$ and $f_{\v{q}}$ (see Fig.~\ref{fig:thermomajorisation}) will consist of linear segments joining the following points:
\begin{subequations}
	\begin{eqnarray}
	f_{\v{p}}&:&\{(0,0),(\gamma_{d-1},p),(\gamma_d+\gamma_{d-1},1)\},\\
	f_{\v{q}}&:&\{(0,0),(\gamma_{d},1-q),(\gamma_d+\gamma_{d-1},1)\}.
	\end{eqnarray}
\end{subequations}

We will now try again to find a candidate state for the join of $\v{p}$ and $\v{q}$. We first note that any state $\v{r}$ that has non-zero occupation in at least two levels that are not the highest-energy levels, say $i$ and $j$, does not thermo-majorise either $\v{p}$ or $\v{q}$. This is because the thermo-majorisation curve of such a state can only reach $1$ at $\gamma_i+\gamma_{j}$, which is bigger than $\gamma_d+\gamma_{d-1}$, and hence $f_{\v{r}}$ will lie above neither $f_{\v{p}}$ nor $f_{\v{q}}$. Let us now focus on the states $\v{r}^{(i)}$ that have a single non-zero entry at $i$-th position for \mbox{$i\in\{1,\dots, d-2\}$}. The thermo-majorisation curve $f_{\v{r}^{(i)}}$ of such a state at a point $\gamma_{d-1}$ will take the value $\gamma_{d-1}/\gamma_{i}<1$. Hence, the curve $f_{\v{r}^{(i)}}$ will not lie above $f_{\v{p}}$ as long as \mbox{$p>\max_i\gamma_{d-1}/\gamma_i=\gamma_{d-1}/\gamma_{d-2}$}. The choice
\begin{equation}
p=\frac{1+\max\left(\frac{\gamma_{d-1}}{\gamma_{d-2}},\frac{\gamma_{d-1}}{\gamma_{d}+\gamma_{d-1}}\right)}{2}
\end{equation}
guarantees this, as well as the consistency with the initial assumption \mbox{$p\geq\frac{\gamma_{d-1}}{\gamma_{d}+\gamma_{d-1}}$}. Thus the only candidate states for the join of $\v{p}$ and $\v{q}$ are of the form $\v{r}=(0,0,\dots,0,r,1-r)$. But, this is exactly a two-level case discussed before and, using the same reasoning, one can show that for any $\v{p}$ and $\v{q}$, consistent with our initial assumptions, the choice of $q$ satisfying
\begin{equation}
\label{eq:ineq}
\frac{\gamma_d}{\gamma_{d-1}}p<(1-q)<\min\left(\frac{\gamma_d}{\gamma_{d-1}},1-\frac{\gamma_{d-1}}{\gamma_d}(1-p)\right),
\end{equation}
guarantees that no join for $\v{p}$ and $\v{q}$ exists. The first inequality on the left guarantees that $f_{\v{q}}$ will be above $f_{\v{p}}$ at the point $\gamma_d$ (refer to Fig.~\ref{fig:thermomajorisation}). Requiring $(1-q)$ to be smaller than the second argument of the minimum guarantees that $f_{\v{q}}$ will be below $f_{\v{p}}$ at the point $\gamma_{d-1}$. Hence, $\v{p}$ and $\v{q}$ are incomparable. Finally, ensuring \mbox{$(1-q)$} to be smaller than the first argument of the minimum guarantees that there exist exactly two candidates for a join of $\v{p}$ and $\v{q}$ and that these are incomparable.

A careful reader would have noticed that in our two-dimensional example we have $\gamma_0=1/2$ not only for infinite temperatures, but also at finite temperatures if the two energy levels are degenerate. More generally, the transformations within any degenerate subspace are governed by the same rules as the infinite-temperature limit from the previous section.\footnote{Note that if some energy states are degenerate, then thermo-majorisation actually forms a preorder. It can be replaced by a partial order only once we identify all states connected via a permutation between degenerate states with corresponding equivalence classes.} Hence, the lattice structure arises within subspaces of states whose energies cannot be distinguished. But degenerate energy subspaces are not the only ones in which the lattice structure can be preserved, as there exist subspaces of $\P_d$ in which thermo-majorisation is effectively described by majorisation. These consist of probability vectors that belong to the same $\beta$-ordering. To see this, consider two probability vectors $\v{p}$ and $\v{q}$ with the same $\beta$-ordering that we will denote by $\beta_1$. Let us also denote by $\v{\gamma}^{\beta_1}$ the \mbox{$\beta_1$-ordered} version of the thermal Gibbs state $\v{\gamma}$. Now, the extremal points of the segments that constitute thermo-majorisation curves $f_{\v{p}}$ and $f_{\v{q}}$ will have the same $x$ coordinates: $x_i=\sum_{j=1}^{i}\gamma_j^{\beta_1}$. Hence, to verify if one curve lies above the other, one only needs to compare their $y$ coordinates. This means that $\v{p}$ thermo-majorises $\v{q}$ if and only if $\v{p}^{\beta_1}\succ\v{q}^{\beta_1}$, where $\succ$ denotes standard majorisation. This allows us to use a slightly modified version of the construction presented in Sec.~\ref{sec:infinite_T_lattice} to find the meet and join. Meet is given by Eq.~\eqref{eq:majo_meet} simply by replacing the entries of $\v{p}$ arranged in a nonincreasing order with the $\beta_1$-ordered entries (and similarly for $\v{q}$). To verify that the resulting probability vector is $\beta_1$-ordered, note that its thermo-majorisation curve is concave. To find the join we also replace nonincreasing order with $\beta_1$-ordering in Eq.~\eqref{eq:majo_join1}, and modify the described iterative procedure in the following way. We define $n$ as the smallest integer such that \mbox{$g_n^{(k)}/\gamma_n^{\beta_1}>g_{n-1}^{(k)}/\gamma_{n-1}^{\beta_1}$}, and by $m$ the greatest integer satisfying
\begin{equation}
\label{eq:thermomajo_join}
\frac{g_{m-1}^{(k)}}{\gamma_{m-1}^{\beta_1}}\geq\frac{\sum_{i=m}^n g_i^{(k)}}{\sum_{i=m}^n \gamma_{i}^{\beta_1}}=:b_k.
\end{equation}
Finally, we define $\v{g}^{(k+1)}$ by setting its components $g_i^{(k+1)}=b_k \gamma_i^{\beta_1}$ for $i\in\{m\dots n\}$ and $g_i^{(k+1)}=g_i^{(k)}$ otherwise. The role of this modified procedure is the same as in Sec.~\ref{sec:infinite_T_lattice}: to ensure that the thermo-majorisation curve of the resulting join $\v{p}\vee\v{q}$ is concave, which also guarantees that $\v{p}\vee\v{q}$ is $\beta_1$-ordered. We thus conclude that if one only considers a subset of classical states that belong to the same $\beta$-ordering, then thermo-majorisation forms a lattice. However, within the full state space it is not the case, since for a given $\v{p}$ and $\v{q}$ there may be multiple incomparable candidates for the join and meet, each belonging to a different $\beta$-ordering (as in Fig.~\ref{fig:thermomajorisation}, where the two dashed lines correspond to two candidates for the join of $\v{p}$ and $\v{q}$).

As a final remark, let us notice that in the infinite-temperature limit states connected by a permutation are reversibly interconvertible, and so they belong to the same thermal equivalence class. Hence, when we speak of a join $\v{r}=\v{p}\vee\v{q}$, it is unique only because states $\v{r}$ and $\Pi\v{r}$, where $\Pi$ denotes arbitrary permutation, are equivalent. At finite temperature, however, this permutation invariance is broken and so is the the uniqueness of join and meet. It is preserved only if we restrict our considerations to a particular class of states described by the same $\beta$-ordering.

\section{Ordering of quantum states}

Let us now proceed to analysing the structure of the thermodynamic ordering of quantum states. In the infinite-temperature limit we will show that, similarly to the classical case, we are dealing with a lattice structure. This could be expected as in this limit the unitary operations are GP and, since unitaries are reversible, for every state with coherence there exists a diagonal (classical) state belonging to the same thermodynamic equivalence class. Therefore, the quantum and classical states share the same structure of thermodynamic ordering.

The situation becomes much more complicated at finite temperatures. In fact, the set of states that a given state $\rho\in\Sc_d$ can be mapped to via GP maps has not been, until recently\footnote{Recently, we became aware of the paper by Buscemi and Gour which provides the necessary and sufficient conditions for the existence of a GP quantum channel between two qubit states in terms of max-relative entropies~\cite{buscemi2017quantum}.}, explicitly found for any dimension~$d$. Therefore, we solve this problem in the simplest case of $d=2$ and provide future thermal cones $\T_+(\rho)$ for all states $\rho\in\Sc_2$. This will allow us to prove that the thermodynamic arrow of time for qubit systems exhibits a lattice structure. This contrast with a classical two-level system provides evidence that coherence may play an important role in thermodynamics by adding structure to the thermodynamic ordering of states. However, whether the lattice structure persists beyond the qubit case for $d\geq 3$ remains an open question for future investigation.

\subsection{Infinite temperature and a lattice structure}

In the infinite-temperature limit the Gibbs state is the maximally mixed state $\gamma=\iden/d$. Hence, GP maps are replaced by unital maps $\E(\iden)=\iden$. The existence of a unital map $\E$ satisfying $\E(\rho)=\sigma$ is equivalent to the spectrum of $\rho$ majorising the spectrum of $\sigma$. To see this, first assume that $\mathrm{spec}(\rho)\succ\mathrm{spec}(\sigma)$. Then, due to the fact that the set of unital maps contains all unitaries, $\rho$ and $\sigma$ can be brought to a diagonal form in the same basis, and the transformation between two diagonal states via a unital CPTP map is described by a bistochastic matrix. Thus, the problem can be mapped to the one discussed in the previous section. Now, assume that neither $\mathrm{spec}(\rho)\succ\mathrm{spec}(\sigma)$ nor $\mathrm{spec}(\sigma)\succ\mathrm{spec}(\rho)$. If there existed a unital channel $\E$ transforming $\rho$ into $\sigma$, then one could also construct a unital channel $\E'$ by composing the conjugate of the unitary diagonalising $\rho$ with $\E$ and with the unitary diagonalising $\sigma$. Such a unital channel $\E'$ would then transform a diagonal state with diagonal $\mathrm{spec}(\rho)$ into a diagonal state with diagonal $\mathrm{spec}(\sigma)$. This would, however, mean that there exists a bistochastic matrix mapping a probability vector $\mathrm{spec}(\rho)$ into $\mathrm{spec}(\sigma)$, which is equivalent to $\mathrm{spec}(\rho)\succ\mathrm{spec}(\sigma)$ and leads to a contradiction with the assumption. We thus conclude that in the infinite-temperature limit majorisation relation between the spectra of two given states is a necessary and sufficient condition for the existence of a GP map between these states.

The slight difference between the quantum and classical scenarios only lies in thermodynamic equivalence classes. Namely, in the classical case these were composed of probability vectors connected via a permutation, whereas in the quantum case these are composed of density matrices connected via a unitary. Regardless of this difference, the partial order between density matrices modulo unitaries forms a lattice.

\subsection{Finite temperatures: qubit evidence for a lattice structure}

We will consider a generic qubit system described by a Hamiltonian $H=E_1\ketbra{E_1}{E_1}$. Let us denote a thermal state of such a system with respect to inverse temperature $\beta$ by $\gamma=e^{-\beta H}/Z$, with $Z=\tr{}{e^{-\beta H}}$. In what follows we will describe qubit states $\rho$, $\rho'$ and a thermal state $\gamma$ using the Bloch sphere representation,
\begin{equation}
\label{eq:bloch_state}
\rho=\frac{\iden+\v{r}_\rho\cdot\v{\sigma}}{2},
\end{equation}
where \mbox{$\v{\sigma}=(\sigma_x,\sigma_y,\sigma_z)$} denotes the vector of Pauli matrices. The Bloch vectors will be parametrised in the following way:
\begin{equation}
\label{eq:bloch_param}
\v{r}_\rho=(x,y,z),\quad\v{r}_{\rho'}=(x',y',z'),\quad\v{r}_\gamma=(0,0,\zeta), 
\end{equation}
where $\zeta=2Z^{-1}-1\geq 0$.

\begin{figure}[t!]
	\begin{centering}	
		\includegraphics[width=\columnwidth]{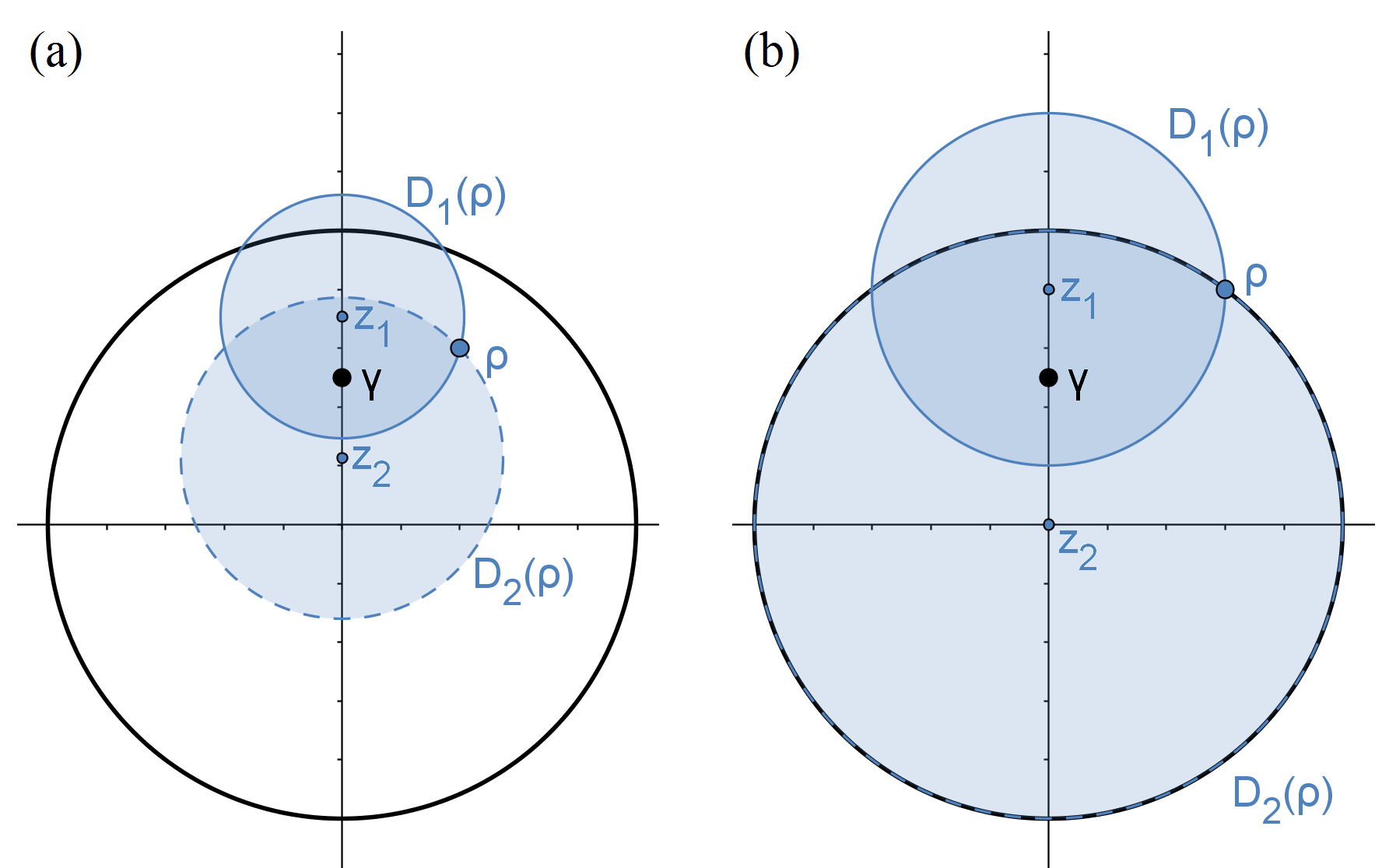}
		\caption[Future thermal cones for qubits (GP operations)]{\label{fig:future_cones}\emph{Future thermal cones for qubits (GP operations).} 
			A general qubit state $\rho$ and a thermal state $\gamma$ with \mbox{$\v{r}_\gamma=(0,0,0.5)$} presented in the Bloch sphere. The disk $D_1(\rho)$ corresponds to a set of states $\{\sigma\}$ with \mbox{$R_-(\sigma)\leq R_-(\rho)$}, whereas the disk $D_2(\rho)$ corresponds to a set $\{\sigma\}$ with \mbox{$R_+(\sigma)\leq R_+(\rho)$}. The equalities are obtained at the edges of the disks, i.e., on circles $C_1(\rho)$ and $C_2(\rho)$. The set of states $\rho$ can be mapped to via GP quantum channels is given by the intersection \mbox{$D_1(\rho)\cap D_2(\rho)$} (which can also be freely revolved around the $z$ axis). (a) A mixed state with \mbox{$\v{r}_\rho=(0.4,0,0.6)$}. (b) A pure state with \mbox{$\v{r}_\rho=(0.6,0,0.8)$}.}
	\end{centering}	
\end{figure}

The following theorem, which may be of independent interest, specifies the necessary and sufficient conditions for the existence of a GP quantum channel between generic qubit states $\rho$ and $\rho'$, provided $\gamma$ is not a pure state (we will comment on this particular zero-temperature case later).
\begin{theorem}[Existence of a GP transformation between qubit states]\label{thm:GP_qubit}
	Consider qubit states $\rho$ and $\rho'$, and the thermal state $\gamma$ that is not pure, i.e., $\zeta\neq 1$. Then, there exists a GP quantum channel $\E$ such that $\E(\rho)=\rho'$ if and only if \mbox{$R_{\pm}(\rho)\geq R_{\pm}(\rho')$}
	where \mbox{$R_{\pm}(\rho)=\delta(\rho)\pm\zeta z$} and
	\begin{equation}
	\label{eq:delta_lattice}
	\delta(\rho):=\sqrt{(z-\zeta)^2+(x^2+y^2)(1-\zeta^2)},
	\end{equation}
	with analogous (primed) definitions for $\rho'$.
\end{theorem}
\noindent The future thermal cone $\T_+(\rho)$ of any qubit state $\rho$ can be found as a corollary of the above theorem.
\begin{corollary}[Future thermal cone of a qubit system]\label{cor:GP_qubit}
	Consider a generic qubit state $\rho$ and orient the Bloch sphere so that its $xz$ plane coincides with the plane containing $\rho$ and a thermal state $\gamma$, i.e., \mbox{$\v{r}_\rho=(x,0,z)$}. Define two disks, $D_1(\rho)$ and $D_2(\rho)$ with corresponding circles $C_1(\rho)$ and $C_2(\rho)$, of radii 	
	\begin{equation}
	\label{eq:radii}
	R_1(\rho)=\frac{R_-(\rho)+\zeta^2}{1-\zeta^2},\quad R_2(\rho)=\frac{R_+(\rho)-\zeta^2}{1-\zeta^2},
	\end{equation}
	centred at 
	\begin{equation}
	\label{eq:centres}
	\begin{array}{ccc}
		\v{z}_1(\rho)&=&[0,0,\zeta(1+R_1(\rho))],\\ \v{z}_2(\rho)&=&[0,0,\zeta(1-R_2(\rho))].
	\end{array}
	\end{equation}
	Then the set of qubit states that a state $\rho$ can be mapped to by GP quantum channels is given, in the Bloch sphere, by the region obtained from revolving the intersection $D_1(\rho)\cap D_2(\rho)$ around the $z$ axis. In other words, it is given by the intersection of two balls of radii $R_1(\rho)$ and $R_2(\rho)$ centred at $\v{z}_1(\rho)$ and $\v{z}_2(\rho)$.
\end{corollary}
\noindent The proof of the above results is based on the Alberti-Uhlmann theorem~\cite{alberti1980problem} and can be found in Appendix~\ref{sec:appendix}. We illustrate the statement of Corollary~\ref{cor:GP_qubit} in Fig.~\ref{fig:future_cones}. Let us also briefly discuss a few particular cases. For every pure state $\rho$ we have \mbox{$x^2+y^2+z^2=1$}, which results in $R_+(\rho)=1$ and $R_-(\rho)=1-2z\zeta$. Hence, pure qubit states are totally ordered by the value of $z$: the top state is described by a Bloch vector $(0,0,-1)$, and the bottom one by $(0,0,1)$. For every incoherent state $\rho$ one of the disks, $D_1(\rho)$ or $D_2(\rho)$, is always contained within the other. Hence, for an incoherent state $\rho$ with $z\geq \zeta$ the future thermal cone is given by $D_1(\rho)$, whereas when $z\leq\zeta$ it is given by $D_2(\rho)$. Finally, note that when $\zeta=0$, i.e., we consider the infinite temperature limit, both disks are centred at the origin and have the same radius equal to the length of $\v{r}_\rho$. We thus recover the majorisation result, as the spectrum of the state $\rho$ majorises that of $\rho'$ if and only if $\v{r}_\rho\geq\v{r}_{\rho'}$.

We also need to comment on the zero-temperature case of pure thermal state $\gamma=\ketbra{E_0}{E_0}$. First, introduce
\begin{equation}
R_3(\rho)=\frac{x^2+(1-z)^2}{2(1-z)},
\end{equation}
and analogously $R_3(\rho')$ for $\rho'$. The conditions of Theorem~\ref{thm:GP_qubit} are then replaced by $z\leq z'$ and $R_3(\rho)\geq R_3(\rho')$. This results in the following change of Corollary~\ref{cor:GP_qubit}. The set of states $\rho'$ that $\rho$ can be transformed via GP maps is given in the Bloch sphere by the intersection of a ball of radius $R_3(\rho)$ centred at \mbox{$\v{z}_3(\rho)=[0,0,1-R_3(\rho)]$} with the half-space defined by $z'\geq z$. The details concerning this special case can also be found in Appendix~\ref{sec:appendix}.

We are now ready to state the anticipated result that the thermodynamic ordering of qubit states at finite temperatures, unlike the ordering of classical states of a two-level system, forms a lattice. Figure~\ref{fig:qubit_lattice} serves as an illustration of the following theorem, the proof of which can be found in Appendix~\ref{sec:appendix2}.

\begin{theorem}[Thermodynamic lattice for qubits]\label{thm:qubit_lattice}
	The thermodynamically ordered set of qubit states forms a lattice. The partially ordered thermodynamic equivalence classes consist of states connected via a unitary conjugation with $U=\exp(-iHt)$. For two distinct equivalence classes consider their representatives, $\rho$ and $\rho'$, living in the $xz$ plane of the Bloch sphere with $x\geq 0$. The join and meet of $\rho$ and $\rho'$  are defined as follows. Introduce 
	\begin{subequations}
		\begin{eqnarray}
		\rho^{\max}_m&=&\arg\max\{R_m(\rho),R_m(\rho')\},\\
		\rho^{\min}_m&=&\arg\min\{R_m(\rho),R_m(\rho')\},
		\end{eqnarray}
	\end{subequations}
	for $m\in\{1,2\}$. The join is then given by a state lying in the Bloch sphere at the intersection of two circles $C_1(\rho^{\max}_1)$ and $C_2(\rho^{\max}_2)$, and the meet is defined analogously by replacing $\max$ with $\min$. 
\end{theorem}

	\begin{figure}[t!]
		\begin{centering}	
			\includegraphics[width=\columnwidth]{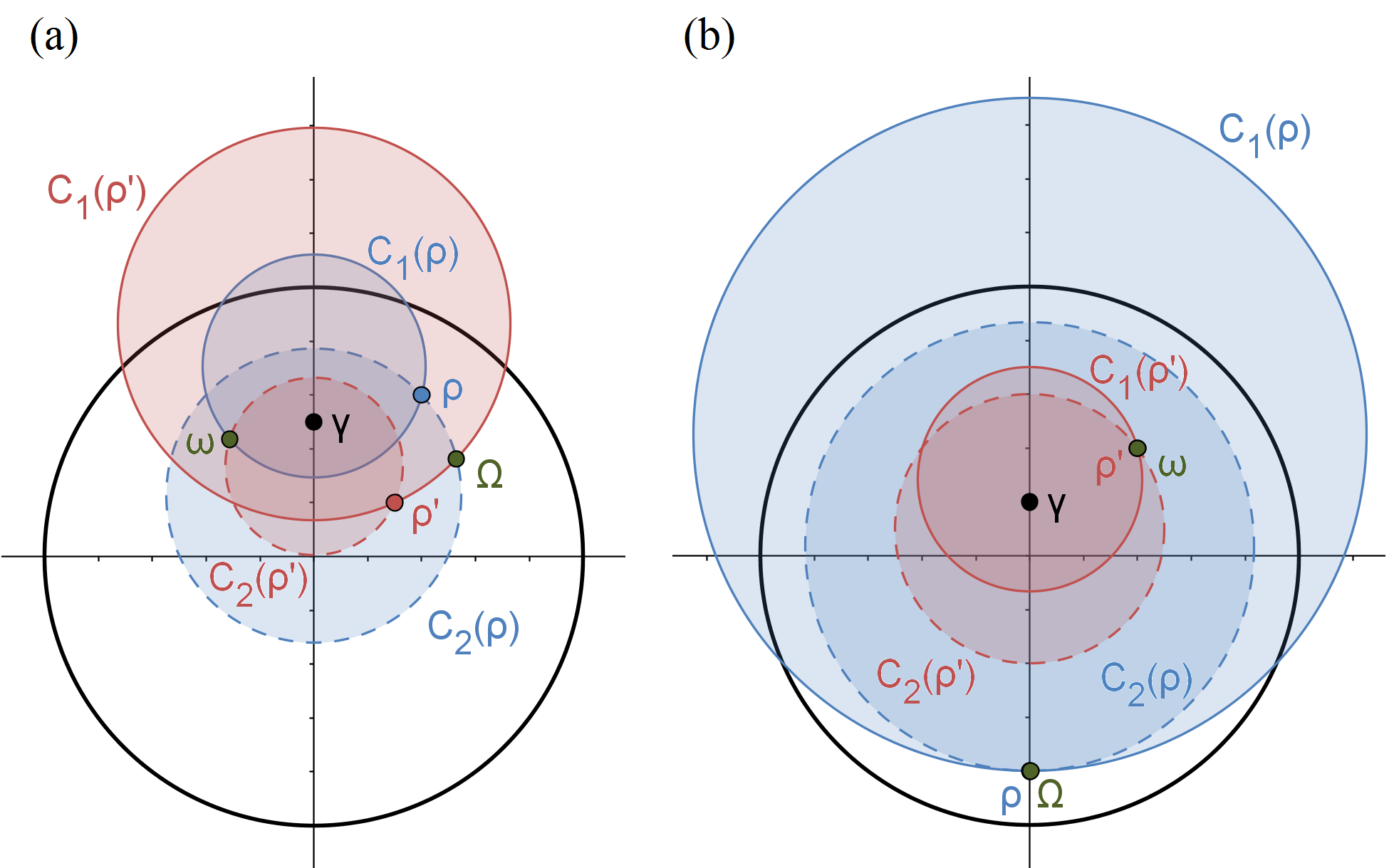}
			\caption[Thermodynamic lattice for qubits]{\label{fig:qubit_lattice}\emph{Thermodynamic lattice for qubits.} A thermal state $\gamma$ and two states $\rho$ and $\rho'$ presented in the Bloch sphere. (a) States described by \mbox{$\v{r}_\rho=(0.4,0,0.6)$}, \mbox{$\v{r}_{\rho'}=(0.3,0,0.2)$}, \mbox{$\v{r}_\gamma=(0,0,0.5)$}. The join $\Omega$ of $\rho$ and $\rho'$ lies at the intersection of $C_1(\rho')$ and $C_2(\rho)$, whereas their meet $\omega$ lies at the intersection of $C_1(\rho)$ and $C_2(\rho')$. (b) States described by \mbox{$\v{r}_\rho=(0,0,-0.8)$}, \mbox{$\v{r}_{\rho'}=(0.4,0,0.4)$}, \mbox{$\v{r}_\gamma=(0,0,0.2)$}. The join $\Omega$ of $\rho$ and $\rho'$ lies at the intersection of $C_1(\rho)$ and $C_2(\rho)$ (which coincides with $\rho$), whereas their meet $\omega$ lies at the intersection of $C_1(\rho')$ and $C_2(\rho')$ (which coincides with $\rho'$).}
		\end{centering}	
	\end{figure}

\section{Interpreting the thermodynamic lattice via the history erasure process}

In the previous two sections we have established the structural properties of the thermodynamic arrow of time and pointed out the differences between quantum and classical scenarios. Here we aim to provide a physical interpretation of these results by introducing and analysing the \emph{history erasure process}. In order to quantitatively analyse this process, we first recall the concept of thermodynamic monotones~\cite{goold2016role}.

Intuitively, it is clear that during evolution along the thermodynamic arrow of time the system must get closer and closer to a thermal Gibbs state or, in over-simplified words, ``entropy has to grow.'' This intuition can be made precise by considering any distance measure $D(\rho,\gamma)$ between a given state $\rho$ and a thermal state $\gamma$ that is contractive under CPTP maps (analogous reasoning applies to classical states for distance measures $D(\v{p},\v{\gamma})$ contractive under stochastic maps). Such measures are known as thermodynamic monotones because they are monotonically non-increasing along the thermodynamic arrow of time. To see this assume that $\E(\rho)=\sigma$ for a GP map $\E$, so that $\sigma$ lies in the future thermal cone of $\rho$. Then,
\begin{equation}
D(\rho,\gamma)\geq D(\E(\rho),\E(\gamma))=D(\sigma,\gamma).
\end{equation}
As a particular example one may consider the relative entropy $S(\rho||\gamma)$ given by
\begin{equation}
S(\rho||\gamma)=\tr{}{\rho\ln\rho}-\tr{}{\rho\ln\gamma},
\end{equation}
which can be interpreted as a non-equilibrium version of the free energy difference between the state $\rho$ and the thermal state $\gamma$~\cite{brandao2013second}. Indeed, we have
\begin{equation}
kTS(\rho||\gamma)=\tr{}{\rho H} - kT S(\rho)+kT\ln Z,
\end{equation}  
which should be compared with the classical expression for the free energy difference $\Delta F=U-TS+kT\ln Z$, with $U$ denoting the average energy. Note that in the infinite temperature limit, $\beta\rightarrow 0$, one has $\gamma=\iden/d$, so that the relative entropy is given by $\ln d -S(\rho)$, and it is the von Neumann entropy that is non-decreasing along the thermodynamic arrow of time. Majorisation and thermo-majorisation conditions can thus be seen as a geometric way of expressing the monotonicity of a whole family of functions that can be interpreted as generalisations of free energy (or entropy in the $\beta\rightarrow 0$ limit). For the clarity of discussion in what follows we will refer mostly to the free energy $S(\rho||\gamma)$ and the von Neumann entropy $S(\rho)$, however, all statements will be valid for any function $D(\rho,\gamma)$ contractive under CPTP maps, unless stated otherwise.

We are now ready to introduce and analyse the history erasure process. Imagine that two possible events may have happened ``in the past'': the system could have been prepared either in the state $\rho$ or in the state $\sigma$. It then evolved along the thermodynamic arrow of time into a state $\tau$, i.e., a GP quantum channel transformed the system state into $\tau$. We now ask: can one infer the past of the system, i.e., whether it was initially prepared in a state $\rho$ or $\sigma$, based on the present state $\tau$? If both $\rho$ and $\sigma$ belong to the past thermal cone $\T_-(\tau)$ then it is impossible, and we say that the $(\rho,\sigma)$-history has been erased during the evolution. Clearly, any history can be erased by evolution that brings the system to the thermal equilibrium state $\gamma$. However, it may not be necessary for the system to evolve all the way to $\gamma$ in order to erase its history. Therefore, the question is as follows: how far along the thermodynamic arrow of time does a system have to evolve for its state to be consistent with both possible pasts, specified by states $\rho$ and $\sigma$? To make the notion of ``far'' precise we may use any thermodynamic monotone; in particular, we will be interested in how much the free energy (entropy) of the system has to decrease (increase) in order to erase its $(\rho,\sigma)$-history.

Let us start from the simplest scenario, when the two possible pasts are thermodynamically ordered. Say that $\rho$ lies in the past thermal cone of $\sigma$. Then, if the system were prepared in the state $\sigma$, it would not need to evolve at all (so its free energy would not need to decrease) in order to achieve history erasure. Indeed, observing the state $\sigma$ one cannot tell whether the system started in $\rho$ and thermalised towards $\sigma$, or if it was prepared in $\sigma$ and did not evolve at all (recall that the identity map is a free thermodynamic operation). A more interesting scenario arises when $\rho$ and $\sigma$ are not ordered. Then, in general, there may be many optimal states $\tau$ which lead to history erasure, i.e., states that lie in the future thermal cones of both $\rho$ and $\sigma$, but whose past thermal cones contain no states with that property. However, if the thermodynamic order has a lattice structure, there is a unique optimal state $\tau$, given by the meet $\rho\wedge\sigma$, that leads to $(\rho,\sigma)$-history erasure. 

The fact that in the infinite temperature limit we deal with a majorisation lattice means that when information is the only thermodynamic resource (in the sense that it does not matter which energy states are occupied, and the only important thing is how ``sharp'' the distribution is), there exists a unique optimal history erasure process. In other words, there is a well-defined way to erase the history while decreasing all thermodynamic monotones (in particular, increasing entropy) in a minimal way. On the other hand, in the classical regime at finite temperatures, when the thermodynamic order is given by thermo-majorisation, there may be many different ways to perform an optimal history erasure process. Recall that in Sec.~\ref{sec:finite_temp} we have seen that already in the simplest case of a two-dimensional distribution there were two candidates for such an optimal state (for the meet of $\v{p}$ and $\v{q}$). This is linked to the fact that at finite temperature it is not only information but also energy that matters, which is reflected by the existence of different $\beta$-orderings. In general, for each of the $d!$ $\beta$-orderings there may be a different candidate for an optimal state. Moreover, the optimal decrease of different thermodynamic monotones may be achieved for different candidate states, so the optimal history erasure process does not exist.

We can also consider a problem dual to history erasure, when instead of erasing two possible pasts we wish to create two possible futures. More precisely, for any two given states ``in the future,'' $\rho$ and $\sigma$, we wish to  find a state $\tau$ that can evolve to both of these states under free thermodynamic operations, i.e., $\T_+(\tau)$ contains both $\rho$ and $\sigma$. The analysis of $(\rho,\sigma)$-future creation process is then analogous to the one presented above, with the optimal state for thermodynamic orders with a lattice structure given by the join $\rho\vee\sigma$ instead of the meet. Alternatively, one may also interpret this dual process as a particular time-reversal of the history erasure. Recall that in Sec.~\ref{sec:partial_order} we introduced two thermodynamic orders: one oriented along the thermodynamic arrow of time, and the other against it. These two orders are mapped into each other by exchanging $\succ$ with $\prec$, i.e., if a state $\tau$ lies in the past thermal cone of both $\rho$ and $\sigma$ according to one order, it will lie in the future thermal cone of both these states according to the dual order. Hence, the optimal state $\tau$ for creating two possible futures, $\rho$ and $\sigma$, is also the optimal state for erasing two possible pasts after reversing the direction of the thermodynamic arrow of time.

Exploiting the properties of the majorisation lattice, we can now prove that in the infinite-temperature limit there is an inherent asymmetry between creating futures and erasing pasts or, in other words, between forward and backward history erasure processes. In Ref.~\cite{cicalese2002supermodularity} it has been shown that the Shannon entropy $H$ is supermodular on a majorisation lattice, meaning that for any two probability distributions $\v{p}$ and $\v{q}$ we have 
\begin{equation}
\label{eq:supermodular}
H(\v{p}\wedge\v{q})+H(\v{p}\vee\v{q})\geq H(\v{p})+H(\v{q}).
\end{equation} 
Note that $\v{p}$ and $\v{q}$ can represent both classical and quantum states, since for $\beta\rightarrow 0$ only the spectrum of a state is important for thermodynamic order. Rearranging the above equation we obtain
\small
\begin{equation}
H(\v{p}\wedge\v{q})-\frac{1}{2}[H(\v{p})+H(\v{q})]\geq \frac{1}{2}[H(\v{p})+H(\v{q})]-H(\v{p}\vee\v{q}).
\end{equation} 
\normalsize
We thus see that the average increase of the Shannon entropy during the optimal $(\v{p},\v{q})$-history erasure (with the average taken over two possible pasts $\v{p}$ and $\v{q}$) is larger than during the optimal $(\v{p},\v{q})$-future creation. We can also say that the average entropy increase during a forward history erasure process is larger than during a backward (time-reversed) history erasure.

Finally, let us discuss the consequences of the existence of a thermodynamic lattice at finite temperatures for qubit systems. First, let us consider two incomparable classical states $\rho$ and $\sigma$. If the state space were restricted only to classical states, there could be two ``optimal'' ways to erase $(\rho,\sigma)$-history, by bringing the system to two incomparable classical states $\v{m}_1$ and $\v{m}_2$ (the candidates for the meet from Sec.~\ref{sec:finite_temp}). However, neither of the two ways would be truly optimal, since some thermodynamic monotones could decrease optimally for $\v{m}_1$, whereas others for $\v{m}_2$. Now, if we are no longer restricted to classical states, there is an optimal way of erasing $(\rho,\sigma)$-history, given by the meet of $\rho$ and $\sigma$. Note that, from Theorem~\ref{thm:qubit_lattice}, this meet is given by a state with coherence. We thus see that for $d=2$ coherence is necessary for the existence of an optimal history erasure process at finite temperatures. Moreover, for the optimal state $\rho\wedge\sigma$ all thermodynamic monotones will be larger than for $\v{m}_1$ or $\v{m}_2$. Hence, exploiting coherence one can erase the classical history of a system using less free energy. One may then wonder whether, independently of the existence of a thermodynamic lattice for $d>2$, coherence allows one to erase history for a smaller free energy cost.

\section{Outlook}

In this work we have just begun to analyse the structure of the thermodynamic arrow of 
time from the point of view of order theory. Most importantly, we provided evidence for potential structural differences between the thermodynamic ordering of classical and quantum states. However, it is crucial to verify whether the coherence-induced lattice structure is still present beyond the investigated qubit case. This will require employing new tools, as the power of the Alberti-Uhlmann theorem on which we based our results is limited to two-dimensional systems. One could, for example, try to develop a quantum analogue of the embedding procedure that in the classical case allows one to arrive at thermo-majorisation condition starting from majorisation~\cite{brandao2013second}. A recent approach based on quantum relative Lorenz curves seems to be a promising avenue here~\cite{buscemi2017quantum}. 

Alternatively, one could investigate how and why the lattice structure breaks at finite temperatures for classical states. Note that the problem comes from the existence of different $\beta$-orderings, as within a particular $\beta$-ordering the lattice structure is preserved. Moreover, in the analysed qubit case we saw that if an incoherent state $\sigma$ can be reached from an incoherent state $\rho$, then there exists a continuous path of states $\rho(t)$, such that $\rho(t)\succ\rho(t')$ for $t\leq t'$ with $\rho(0)=\rho$ and $\rho(1)=\sigma$. On the other hand, if such a path is restricted to classical states, then it exists only if $\rho$ and $\sigma$ belong to the same $\beta$-ordering. Hence, the lattice structure may arise due to coherence providing the ``continuous connection'' between states belonging to different $\beta$-orderings. Here, linking with the known results concerning relative majorisation for continuous probability distributions may be useful~\cite{van2010renyi}.

Furthermore, in all the cases where the thermodynamic ordering forms a lattice, one can use its structure to find new thermodynamic relations.  For example the authors of Ref.~\cite{cicalese2002supermodularity} have shown that the Shannon entropy
$H$ is supermodular on a majorisation lattice [see Eq.~\eqref{eq:supermodular}] and also subadditive, meaning that \mbox{$H(\v{p}\wedge\v{q})\leq H(\v{p})+H(\v{q})$}. One could then ask whether similar relations hold at finite temperatures when $H$ is replaced by some thermodynamic monotone function, e.g., by the free energy. 

Finally, one may investigate other resource theories from the order-theoretic point of view to get insight into their structure. For example, note that transformations between pure states in both the resource theory of entanglement~\cite{nielsen2010quantum} and coherence~\cite{du2015conditions} are ruled by majorisation partial order, so actually by a majorisation lattice. In fact, very recently this structure was used to study approximate transformations between pure bipartite entangled states~\cite{bosyk2017approximate}. On the other hand, in the resource-theoretic formulation of thermodynamics using thermal operations~\cite{horodecki2013fundamental}, it is known that the past thermal cone of each pure qubit state consists (up to equivalence class) only of that state~\cite{lostaglio2015quantum}. It is therefore impossible to define a join for two distinct pure states and the partial ordering of states defined by thermal operations does not form a lattice.

\bigskip

\textbf{Acknowledgements:} I would like to thank Antony Milne and Matteo Lostaglio for helpful comments and discussions. I am also very grateful for the ongoing support provided by David Jennings and Terry Rudolph. This work was supported by EPSRC and in part by COST Action MP1209. I also acknowledge support from the ARC via the Centre of Excellence in Engineered Quantum Systems (EQuS), project number CE110001013.

\appendix

\section{Proof of Theorem~1}
\label{sec:appendix}

\begin{proof}
The proof of Theorem~\ref{thm:GP_qubit} is based on the Alberti-Uhlmann theorem~\cite{alberti1980problem}. In the qubit case this yields necessary and sufficient conditions for the existence of a CPTP map $\E$ such that $\E(\rho)=\rho'$ and $\E(\sigma)=\sigma'$. These conditions are given by:
\begin{equation}\small
\norm{\lambda \rho-(1-\lambda)\sigma}_1\geq \norm{\lambda \rho'-(1-\lambda)\sigma'}_1 \quad \forall \lambda\in[0,1],
\end{equation}\normalsize
where $\norm{A}_1=\tr{}{\sqrt{AA^\dagger}}$. Thus, by simply setting \mbox{$\sigma=\sigma'=\gamma$}, we obtain necessary and sufficient conditions for the existence of a GP quantum map $\E$ such that \mbox{$\E(\rho)=\rho'$}. Using the fact that norms are non-negative, these can be expressed by
\begin{equation}
\label{eq:Delta_condition}
\Delta_\lambda:=D_\lambda(\rho)- D_\lambda(\rho')\geq 0\quad \forall \lambda\in[0,1],
\end{equation} 
where we have defined \mbox{$D_\lambda(\rho):=\norm{\lambda \rho-(1-\lambda)\gamma}_1^2$}.

Before we find necessary and sufficient conditions for Eq.~\eqref{eq:Delta_condition} to hold, let us first simplify the problem. Namely, note that states connected via a unitary \mbox{$U(t)=e^{-iHt}$} are reversibly interconvertible under GP operations, and thus belong to the same equivalence class. Hence, we can focus only on one representative of this class lying in the $xz$ plane of the Bloch sphere with $x\geq 0$. This means that instead of considering general Bloch vectors of the form $\v{r}_\rho=(x,y,z)$ we can focus only on the ones given by $\v{r}_\rho=(x,0,z)$.

Now, using the parametrisation of qubit states introduced in Eqs.~\eqref{eq:bloch_state}-\eqref{eq:bloch_param}, we can write
\begin{equation}
\label{eq:D_square}
D^2_\lambda(\rho)=a(\rho)+\sqrt{a(\rho)^2-b(\rho)^2},
\end{equation}
where
\small
	\begin{eqnarray*}
	a(\rho)&=&\left(2+\frac{x^2+(z+\zeta)^2}{2}\right)\lambda^2-\left(2+\zeta(z+\zeta)\right)\lambda+\frac{1+\zeta^2}{2},\\
	b(\rho)&=&(1 - 2 \lambda) \sqrt{\zeta^2 - 
		2 \zeta (z + \zeta) \lambda + (x^2 + (z + \zeta)^2)\lambda^2}.
	\end{eqnarray*}\normalsize
However, notice that
\begin{equation*}
a(\rho)^2-b(\rho)^2=\left[a(\rho)-(1-2\lambda)^2\right]^2=:c(\rho)^2,
\end{equation*}
which means that
\begin{equation}
D^2_\lambda(\rho)=\left\{
\begin{array}{lll}
(1-2\lambda)^2&:&c(\rho)\leq 0,\\
(1-2\lambda)^2+2c(\rho)&:&c(\rho)>0.
\end{array}
\right.
\end{equation}
Since $c(\rho)$ is quadratic in $\lambda$, one can explicitly express the regions of $\lambda$ with different solutions for $D_\lambda^2(\rho)$ by finding the zeros of \mbox{$c(\rho)=A(\lambda-\lambda_1)(\lambda-\lambda_2)$}. Using elementary calculus one can show by direct calculation that \mbox{$0\leq\lambda_1\leq\frac{1}{2}\leq\lambda_2\leq 1$} and \mbox{$A\leq 0$}. Hence, we can rewrite Eq.~\eqref{eq:D_square} as  
\begin{equation*}
D^2_\lambda(\rho)=\left\{
\begin{array}{lll}
(1-2\lambda)^2&:&\lambda\in[0,\lambda_1]\mathrm{~and~}\lambda\in[\lambda_2,1],\\
(1-2\lambda)^2+2c(\rho)&:&\lambda\in(\lambda_1,\lambda_2).
\end{array}
\right.
\end{equation*}
We also note that for $\zeta=1$ (when the Gibbs state is pure) we have $\lambda_1=0$ independently of the state $\rho$, so instead of three there are only two different regions of $\lambda$ with different solutions for $D_\lambda^2(\rho)$.

\begin{figure}[t!]
	\begin{centering}
		\includegraphics[width=0.9\columnwidth]{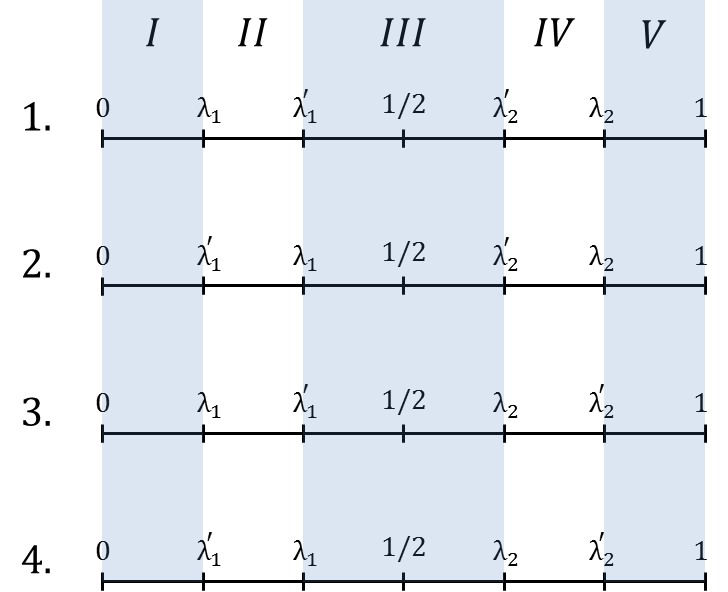}
		\caption[Possible orderings of eigenvalues]{\label{fig:eigen_order}\emph{Possible orderings of eigenvalues.} The orderings 1--4 of eigenvalues $\{\lambda_1,\lambda_1',\lambda_2,\lambda_2'\}$, together with labels I-V corresponding to different regions of the $[0,1]$ interval.}
	\end{centering}
\end{figure}

By analogously introducing $c(\rho')$ with solutions given by $\lambda_1'$ and $\lambda_2'$, the necessary and sufficient conditions for the existence of a GP map $\E$ such that $\E(\rho)=\rho'$ [specified by Eq.~\eqref{eq:Delta_condition}] can be expressed as follows. There are four ways the eigenvalues \mbox{$\{\lambda_1,\lambda_1',\lambda_2,\lambda_2'\}$} can be ordered and these are depicted in Fig.~\ref{fig:eigen_order} (assuming $\zeta\neq 1$; for $\zeta=1$ case see below). In regions I and V (see Fig.~\ref{fig:eigen_order}), independently of the ordering, we have \mbox{$\Delta_\lambda=0$}. Now, in region II we have \mbox{$\Delta_\lambda=2c(\rho)\geq 0$} for orderings 1. and 3., whereas for orderings 2. and 4. we have \mbox{$\Delta_\lambda=-2c(\rho')\leq 0$}. Similarly, in region IV one has that $\Delta_\lambda$ is equal to \mbox{$2c(\rho)\geq 0$} for orderings 1. and 2., whereas for orderings 3. and 4. it is equal to \mbox{$-2c(\rho')\leq 0$}. Therefore, for $\Delta_\lambda$ to be positive for all \mbox{$\lambda\in[0,1]$}, the eigenvalues must be ordered according to ordering 1. The remaining condition to check is whether within region III this ordering also results in $\Delta_\lambda\geq 0$. We have \mbox{$\Delta_\lambda=2[c(\rho)-c(\rho')]$} and since both quadratic functions, $c(\rho)$ and $c(\rho')$, share the same constant term and $\lambda\neq 0$, the problem can actually be simplified to comparing two linear functions. This, in turn, can be done by simply comparing the functions at the edge of the region III, where  $c(\rho)\geq 0$ and $c(\rho')=0$. Therefore, we conclude that Eq.~\eqref{eq:Delta_condition} holds if and only if $\lambda_1\leq\lambda_1'$ and $\lambda_2\geq\lambda_2'$. For the case of $\zeta=1$ the analysis is similar: one still needs $\lambda_2\geq\lambda_2'$, however since $\lambda_1=\lambda_1'=0$, one needs to verify that the derivative of $\Delta_\lambda$ over $\lambda$ at $\lambda=0$ is positive. This results in additional condition that $z\leq z'$.

In the final step of the proof we need to use explicit expressions for $\lambda_1$ and $\lambda_2$,
\begin{subequations}
	\begin{eqnarray}
\label{eq:lambdas}
\lambda_1&=&\frac{2 - \zeta (z + \zeta) - \delta}{4-(z +\zeta)^2-x^2},\\
\lambda_2&=&\frac{2 - \zeta (z + \zeta) + \delta}{4-(z +\zeta)^2-x^2},
	\end{eqnarray} 
\end{subequations}
where $\delta$ is given by Eq.~\eqref{eq:delta_lattice}. By solving the above equations for $z$ with fixed $\lambda_m$ (with $m\in\{1,2\}$), one can find that the region of fixed $\lambda_m$ is given by a circle centred at $\v{z}_m=[0,0,\zeta(\lambda_m^{-1}-1)]$ and of radius \mbox{$R_m=|\lambda_m^{-1}-2|$}. It is a straightforward calculation to show that for $\zeta\neq 1$ these centres and radii correspond exactly to the ones stated in Corollary~\ref{cor:GP_qubit}. For $\zeta=1$ the condition coming from $\lambda_1$ has already been included by $z\leq z'$, and the one coming from $\lambda_2$ corresponds to a circle of radius $R_3$ centred at $\v{z}_3$ as described in the main text. Moreover, since \mbox{$0\leq\lambda_1\leq\frac{1}{2}\leq\lambda_2\leq 1$}, we have 
\begin{subequations}
	\begin{eqnarray}
	\lambda_1\leq\lambda_1'&\Leftrightarrow&R_1(\rho)\geq R_1(\rho'),\\
	\lambda_2\geq\lambda_2'&\Leftrightarrow&R_2(\rho)\geq R_2(\rho').
	\end{eqnarray} 
\end{subequations}
Hence, the necessary and sufficient condition for the existence of a GP map between $\rho$ and $\rho'$, specified by Eq.~\eqref{eq:Delta_condition}, is that $R_1(\rho)\geq R_1(\rho')$ and $R_2(\rho)\geq R_2(\rho')$. This can be equivalently expressed with the use of simplified variables $R_\pm(\rho)$ as in Theorem~\ref{thm:GP_qubit}. Finally, we note that given two circles of radii $R_1$ and $R_1'$, centred at $(0,0,\zeta(1+R_1))$ and $(0,0,\zeta(1+R_1'))$, respectively, we have that the circle with smaller radius is contained within the circle of bigger radius. The same holds true for circles of radii $R_2$ and $R_2'$ centred at $(0,0,\zeta(1-R_2))$ and $(0,0,\zeta(1-R_2'))$. We have thus finished the proof of Corollary~\ref{cor:GP_qubit}.
\end{proof}

\section{Proof of Theorem~3}
\label{sec:appendix2}

\begin{proof}
	First of all, we can restrict our considerations to states lying in the $xz$ plane of the Bloch sphere with Bloch vectors \mbox{$(x,0,z)$} and \mbox{$(-x,0,z)$} being equivalent. Now, assume that two states, $\rho$ and $\rho'$, are comparable, $\rho\succ\rho'$. Then, by Theorem~\ref{thm:GP_qubit} and Corollary~\ref{cor:GP_qubit}, the two disks $D_1(\rho')$ and $D_2(\rho')$ are fully contained inside $D_1(\rho)$ and $D_2(\rho)$. As a result \mbox{$\rho_1^{\max}=\rho$}, \mbox{$\rho_2^{\max}=\rho$}, \mbox{$\rho_1^{\min}=\rho'$} and \mbox{$\rho_2^{\min}=\rho'$}. Hence, the join is given by $\rho$ and the meet by $\rho'$, consistent with the fact that for every lattice if $\rho\succ\rho'$ then \mbox{$\rho\vee\rho'=\rho$} and \mbox{$\rho\wedge\rho'=\rho'$}.
	
	Now, let us consider the case when $\rho$ and $\rho'$ are incomparable, i.e., neither $\rho\succ\rho'$ nor $\rho'\succ\rho$. Then without loss of generality we have
	\mbox{$R_+(\rho)>R_+(\rho')$} and \mbox{$R_-(\rho)<R_-(\rho')$}. Consider a set states $\T_-(\rho,\rho')$ whose future thermal cones contain both $\rho$ and $\rho'$. According to Theorem~\ref{thm:GP_qubit}, \mbox{$\tau\in\T_-(\rho,\rho')$} if and only if \mbox{$R_+(\tau)\geq R_+(\rho)$} and \mbox{$R_-(\tau)\geq R_-(\rho')$}. Now, if there existed a state $\Omega$ such that \mbox{$R_+(\Omega)=R_+(\rho)$} and \mbox{$R_-(\Omega)=R_-(\rho')$} it would clearly be a join \mbox{$\rho\vee\rho'$}. This is because one could reach both $\rho$ and $\rho'$ from $\Omega$ and also $\Omega$ itself could be reached from any \mbox{$\tau\in\T_-(\rho,\rho')$}. We will now prove that such a state $\Omega$ exists for any choice of incomparable states $\rho$ and $\rho'$. The condition \mbox{$R_+(\Omega)=R_+(\rho)$} means that \mbox{$\Omega\in C_2(\rho)$}, whereas the condition \mbox{$R_-(\Omega)=R_-(\rho')$} means that \mbox{$\Omega\in C_1(\rho')$}. Hence, such a state exists if and only if the circles $C_2(\rho)$ and $C_1(\rho')$ intersect (see Fig.~\ref{fig:qubit_lattice}). To prove this first note that a thermal state $\gamma$ is contained inside both circles [straightforward from Eqs.~\eqref{eq:radii}-\eqref{eq:centres}]. This means that either the circles intersect or one is fully contained inside the other. However, the latter is not possible, because a circle $C_2(\rho)$ contains a point $\rho$ that is inside $C_1(\rho')$; and a circle $C_1(\rho')$ contains a point $\rho'$ that is inside $C_2(\rho)$. We thus conclude that the circles $C_2(\rho)$ and $C_1(\rho')$ do intersect and that a state $\Omega$ lying at their intersection is a join \mbox{$\rho\vee\rho'$}.
	
	Analogously, if a state $\omega$ exists such that \mbox{$R_+(\omega)=R_+(\rho')$} and \mbox{$R_-(\omega)=R_-(\rho)$}, then it would be a meet \mbox{$\rho\wedge\rho'$}. The existence of such state is equivalent to the circles $C_1(\rho)$ and $C_2(\rho')$ intersecting. Again, the state $\gamma$ is contained in both circles, so that they either intersect or one is contained inside the other. The latter is impossible, because  a circle $C_1(\rho)$ contains a point $\rho$ that is outside $C_2(\rho')$; and a circle $C_2(\rho')$ contains a point $\rho'$ that is outside $C_1(\rho)$. Hence, the circles $C_1(\rho)$ and $C_2(\rho')$ do intersect and that a state $\omega$ lying at their intersection is a meet \mbox{$\rho\wedge\rho'$}.
	
\end{proof}

\bibliographystyle{apsrev4-1}
\bibliography{Bibliography_thermodynamics,Bibliography_uncertainty}

\end{document}